\documentclass[prb, superscriptaddress, floatfix,twocolumn,showpacs,amsfonts,amsmath,amssymb,final]{revtex4}

\usepackage{graphicx}

\usepackage{epstopdf}
\usepackage[countmax]{subfloat}

\usepackage{dcolumn} % align table columns on decimal point

\usepackage{bm} % bold math

\usepackage{color}

\usepackage{float}

\usepackage{latexsym}

\usepackage{feynmf}

\begin{document}

\title{Spin liquid phase in the semiclassical theory of the Heisenberg model on an anisotropic triangular lattice with ring exchange}

\author{Michael Holt}

\affiliation{Centre for Organic Photonics and Electronics, School of Mathematics \& Physics, University of Queensland, Brisbane, Queensland 4072, Australia}

\author{Ben J. Powell}

\affiliation{Centre for Organic Photonics and Electronics, School of Mathematics \& Physics, University of Queensland, Brisbane, Queensland 4072, Australia}

\author{Jaime Merino}

\affiliation{Departamento de F\'isica Te\'orica de la Materia Condensada, Condensed Matter Physics Center (IFIMAC) and Instituto Nicol\'as Cabrera, Universidad Aut\'onoma de Madrid, Madrid 28049, Spain}

\date{\rm\today}

\begin{abstract}

{We investigate the effect of ring-exchange on the ground-state properties and magnetic excitations of the $S = 1/2$ Heisenberg model on the anisotropic triangular lattice with ring-exchange at $T = 0$ using linear spin-wave theory. Classically, we find stable N\'{e}el, spiral and collinear magnetically ordered phases. Upon including quantum fluctuations to the model, linear spin-wave theory shows that ring exchange induces a large quantum disordered region in the phase diagram, completely wiping out the classically stable collinear phase. Analysis of the spin-wave spectra for each of these three models demonstrates that the large spin-liquid phase observed in the full model is a direct manifestation of competing classical orders. To understand the origin of these competing phases we introduce models where either the four spin contributions from ring exchange, or the renormalization of the Heisenberg terms due to ring exchange are neglected. We find that these two terms favor rather different physics.}

\end{abstract}


% 75.10.Jm Quantized spin models

% 75.10.Kt Spin Liquids, Valence Bond phases, and related phenomena

% 75.30.Ds Spin Waves

% 75.50.Ee Antiferromagnetics 

\maketitle

\section{Introduction}

\label{sec:Introduction}

Quantum spin liquids are characterized by ground states with no long-range magnetic order and no breaking of spatial (rotational or translational) symmetries that are not adiabatically connected to the band insulator \cite{Balents10, Normand09, Powell11}. Recently a number of experiments have identified a handful of materials as candidate spin liquids \cite{Lee08, Shimizu03, Yamashita08, Yamashita09, Yamashita10, Itou10, Kanoda10,Fjaerestad07,Shirata12,Zhou11,Shimizu07}. Both the organic charge transfer salts $\kappa$-(BEDT-TTF)$_2$Cu$_2$(CN)$_3$ (Ref. \onlinecite{Shimizu03, Yamashita08, Yamashita09, Kanoda10}) and Me$_3$EtSb[Pd(dmit)$_2$]$_2$ (Ref. \onlinecite{Yamashita10, Itou10, Kanoda10}), where Et =  C$_{2}$H$_{5}$ and Me =  CH$_3$ are spin-liquid candidates. However, other members of the organic charge transfer salt families $\kappa$-(BEDT-TTF)$_2X$ and $Y$[Pd(dmit)$_2$]$_2$ display long range magnetic order, for example, $X=$Cu[N(CN)$_2$]Cl or Cu[N(CN)$_2$]Br (for deuterated BEDT-TTF) and $Y=$Me$_4$P, Me$_4$As, EtMe$_3$As, Et$_2$Me$_2$P,  Et$_2$Me$_2$As and Me$_4$Sb.\cite{Shimizu07, Powell11, Kanoda10} Additionally, the inorganic materials Cs$_2$CuCl$_4$ \cite{Fjaerestad07},  Ba$_3$CoSb$_2$O$_{9}$ \cite{Shirata12} and  Ba$_3$CuSb$_2$O$_{9}$ \cite{Zhou11} have also been suggested to be spin-liquid candidates.

\begin{figure}
 \begin{center}
     \includegraphics[width=0.95\columnwidth, trim = 0mm 40mm 0mm 40mm, clip]{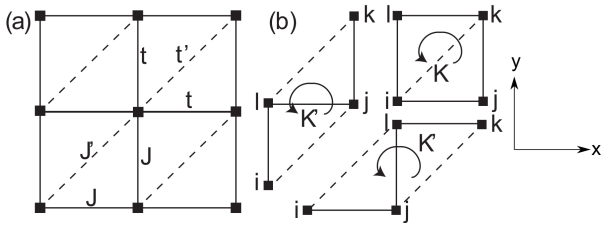}
    \caption{(a) Sketch of the anisotropic triangular lattice showing the exchange interactions $J$ and $J'$, which act in the nearest-neighbor and next-nearest neighbor along one diagonal directions as shown.  (b) The three distinct ways to draw the four-site plaquettes relevant to ring exchange on the anisotropic triangular lattice. We have also shown the $x$-$y$ coordinate system in which we perform our analysis}
    \label{fig:schematic}
    \end{center}
\end{figure}

The simplest model for the Mott insulating states of the $\kappa$-(BEDT-TTF)$_2X$ and $Y$[Pd(dmit)$_2$]$_2$ salts is the half-filled Hubbard model on the anisotropic triangular lattice \cite{Powell11} (see Fig. \ref{fig:schematic}(a)), where each site represents a dimer, (BEDT-TTF)$_2$ or [Pd(dmit)$_2$]$_2$. This model contains three parameters: $U$ the effective on-site Coulomb repulsion, $t$ the nearest neighbor hopping integral and $t'$ the next-nearest neighbor hopping integral along one diagonal only. The Hubbard model on the anisotropic triangular lattice has been studied via a number of approaches \cite{RVB3, Watanabe06, Sahebsara06, Kyung06, RVB, Chen13, Tocchio13}. Some methods have suggested that a spin liquid is realised in the insulating phase.

For $U\gg t,t'$, i.e., deep in the Mott insulating phase, the model simplifies further to the Heisenberg model on the anisotropic triangular lattice with $J=4t^2/U$ and $J'=4t'^2/U$ to leading order. Electronic structure calculations on the anisotropic triangular lattice \cite{Scriven12, Kandpal09, Nakamura09, Nakamura12, Tsumuraya13} suggest that both spin liquids $\kappa$-(BEDT-TTF)$_2$Cu$_2$(CN)$_3$ and Me$_3$EtSb[Pd(dmit)$_2$]$_2$  and the valence bond-solid, Me$_3$EtP[Pd(dmit)$_2$]$_2$, have $0.5\lesssim J'/J\lesssim0.8$; whereas salts that display long range order have either  $J'/J\lesssim0.5$ or $J'/J\gtrsim 0.8$. The anisotropic triangular lattice is also realized in Cs$_2$CuBr$_4$ ($J'/J\approx 2$) and Cs$_2$CuCl$_4$  ($J'/J\approx 3$) \cite{Fjaerestad07} and the isotropic limit ($J'/J = 1$) describes Ba$_3$CoSb$_2$O$_{9}$ and  Ba$_3$CuSb$_2$O$_{9}$ \cite{Susuki13}. 

Many of the organic charge transfer salts considered here undergo Mott metal-to-insulator transitions under relatively modest hydrostatic pressures \cite{ Kanoda10, Powell06}. This suggests that higher order terms in the $U/t$ expansion may be relevant. Furthermore, there is significant variation in the critical pressure required to drive the Mott transition in different salts \cite{Yamaura04} which suggests that different salts represent different values of $U/t$ and not just different values of $t'/t$. There has been far less investigation of how $U/t$ affects the properties of the materials than $t'/t$. If one continues to integrate out the charge degrees of freedom, the first non-trivial new terms appear at fourth order with the `ring-exchange' processes illustrated in Fig. \ref{fig:schematic}(b) [see also Eq. (\ref{eq:bigHamiltonian}), below]. Such ring exchange processes frustrate the system. There are two distinct ring exchange terms on the anisotropic triangular lattice $K=80t^4/U^3$ and $K'=80t^2t'^2/U^3$ to lowest order \cite{MacDonald88, Balents03}, which originate from the different ways to arrange four-sites on the anisotropic triangular lattice. Note that the large prefactor means that the ring exchange term is relevant to larger values of $U/t$ than one would expect na\"ively. It has been argued \cite{Motrunich05, Yang10} that near the Mott transition ring exchange destroys the long range magnetic order. In particular, for $J'=J$ and $K'=K$ Motrunich \cite{Motrunich05} found that AFM order is preserved for small $K/J \lesssim 0.14 - 0.20$ \cite{fn1} but is destroyed for larger $K/J$ leading to a gapped spin liquid for $K/J > 0.28$. However, this implies that applying pressure, which decreases $U/t$, should drive a magnetically ordered to spin liquid transition; which has not been observed in the antiferromagnetically ordered organic charge transfer salts with $t'\simeq t$.

The isotropic triangular lattice with multiple-spin exchange has been widely studied since the 1960s in the context of solid $^3$He; for an extensive review of magnetism in solid $^3$He see Roger \cite{Roger83} and references therein. Since the early studies by Thouless,\cite{Thouless65} single monolayers of solid $^3$He have been absorbed on graphite \cite{Franco86, Godfrin88} and several studies were performed to develop a theoretical understanding of the experiments in terms of multiple spin-exchange models \cite{Kubo97, Momoi97, Kubo98, Misguich98, Misguich99, Roger98, Momoi99, Kubo03, Yasuda06}. Additionally, multiple spin-exchange models have gained interest in frustrated spin systems, in particular the parent cuprate high-temperature superconductors \cite{Sugai90, Coldea01, Nunner02}.

The Heisenberg model on the isotropic triangular lattice without ring exchange ($J' = J$, $K = K' = 0$) has been studied extensively \cite{Balents10, Powell11}.  Anderson first proposed the resonating valence bond (RVB) spin liquid state as a possible ground state of the isotropic triangular lattice \cite{Anderson74}. However, later numerical work \cite{Sindzingre94} has shown that the ground state has `$120^\circ$ order' - a special case of  spiral order, discussed below, with an ordering wavevector ${\bf Q}=(2\pi/3, 2\pi/3)$. A range of other methods have been used to study the Heisenberg model on the anisotropic triangular lattice including linear spin wave theory \cite{Merino99, Trumper99}, modified spin-wave theory \cite{Hauke13}, series expansions \cite{Fjaerestad07, Zheng99}, the coupled cluster method \cite{Bishop09}, large-N expansions \cite{ChungJPCM01}, variational Monte Carlo \cite{Monte}, resonating valence bond theory \cite{RVB, RVB1, RVB2, RVB3, RVB4}, pseudo-fermion functional renormalization group \cite{Reuther11}, slave rotor theory \cite{Rau11}, renormalisation group \cite{Starykh07}, and the density matrix renormalisation group \cite{Weng06}. These calculations show that for small $J'/J$, N\'eel $(\pi, \pi)$ order is realised and spiral $(q, q)$ long range AFM order is realized for $J'/J\sim1$. There remains controversy as to whether another state is realized between these two phases, but no conclusive evidence for a spin liquid ground state has been found in this model. 

The only works we are aware of to discuss the Heisenberg model on the anisotropic triangular lattice with ring exchange consider two leg \cite{Sheng09} and four leg \cite{Block11} ladders. Both studies suggest the existence of quantum spin liquids. Therefore, it is important to ask how these states survive as one moves to the full two-dimensional problem. 

Hauke \cite{Hauke13} has considered the fully anisotropic triangular lattice and argued that this can explain the phase diagram of the organic charge transfer salts. Alternative models such as the quarter-filled Hubbard model, with each site representing a monomer \cite{Li10} and multi-orbital models \cite{Nakamura12} have also been proposed. These works do not consider ring exchange and it is believed that additional interaction terms beyond the nearest neighbour Heisenberg model need to be included to help understand and explain why some materials have magnetically ordered or spin liquid ground states.

Very recently we studied the effect of third nearest-neighbour interactions within a mean-field Schwinger-boson framework \cite{Merino14} and found a spin-liquid phase for $J'/J > 1.8$ and $J_{3}/J \lesssim 0.1$. Such terms can arrive as the two-spin contribution from ring exchange. Majumdar {\it et al}. \cite{Majumdar12} has performed a spin-wave theory study to order $1/S^2$ of the effect of ring exchange on the N\'{e}el phase, but only considered the four-spin terms and neglected the two-spin renormalization. It is therefore interesting to study the effect of the full ring-exchange term on the observed magnetic properties of the anisotropic triangular lattice.

The aim of the present work is to investigate the effect of ring exchange on the magnetic properties of the anisotropic triangular lattice, using linear spin-wave theory. Linear spin-wave theory provides an important benchmark, and is a good starting point for more technical studies. Furthermore, we will explicitly compare the behaviour of the model keeping only spin-exchange or four-spin exchange terms from ring exchange with the full model considering both exchange terms.  

For the full model we found N\'{e}el order is robust to ring exchange for $J'/J \ll 1$, being stable up to around $J'/J \approx 0.59$ for $K/J \approx 0.12$. For $K/J > 0.12$, minima develop along the $(k,k)$ direction which cause the N\'{e}el order to become destabilized. We also found that the spiral phase was dramatically suppressed in the quantum calculations. With strong quantum fluctuations, the spiral phase was only able to survive up to $K/J = 0.10$ for $J'/J = 1$ in constast to $K/J = 1/3$ found classically. In the weakly-coupled chain limit of our model $J'/J \gg 1$ we found spiral order persists in the presence of weak frustration from ring exchange ($K'/K \ll 1$). This is highly analogous to the `order-by-disorder' mechanism due to quantum or thermal fluctuations \cite{Villain, Chandra90, Chubukov92}. In a large region of the quantum phase diagram, the classically stable collinear phase is wiped out and replaced with a spin-liquid phase. Analysis of the spin-wave spectra show that the spin liquid is a consequence of competition between classical ordered states. 

%Considering the relative effects of each of the ring exchange contributions (the two-spin and four-spin exchange terms), we found that their preferred ground-states compete, resulting in a large spin-liquid region in the full model. Thus we conclude that the interplay of the ring exchange terms is responsible for the spin-liquid state found in the full model. Our results are relevant to weak Mott insulators, i.e., insulators lying close to the insulator-to-metal transition so that ring-exchange is relevant.

The present work is organised as follows: in section \ref{sec:Model} we introduce the anisotropic triangular lattice with ring exchange model. We consider the classical phase diagrams for three variants of the ring exchange model \ref{sec:Classical}. In section \ref{sec:LSWT} we present the spin-wave theory formalism, while in section \ref{sec:GroundState} we consider the ground-state properties: quantum phase diagram and staggered magnetization of the three models studied. In section \ref{sec:Excitations} we discuss the elementary excitations of the three models, in particular (i) the existence of minima along the diagonal in the N\'{e}el phase and (ii) why the collinear phase is so fragile to quantum fluctuations. In section \ref{sec:Organics} we relate our findings to the organic materials.  Finally, in section \ref{sec:Conclusions} we present our conclusions.

\section{Heisenberg Model on an anisotropic triangular lattice with ring exchange}

\label{sec:Model}

We are interested in understanding the magnetic properties of the $S = 1/2$ multiple-spin exchange Hamiltonian \cite{Thouless65, Misguich99} on an anisotropic triangular lattice involving ring exchange on four sites at $T = 0$:
\begin{eqnarray}
\label{eq:bigHamiltonian}
\hat{H} &=& \frac{J}{2}\sum_{
\begin{picture}(11,11)(0,0)
	\put (0,9) {\line (1,0) {10}}
	\put (0,9) {\circle*{4}}
	\put (10,9) {\circle*{4}}
\end{picture}}\hat{P}_{ij} +  \frac{J}{2}\sum_{
\begin{picture}(11,11)(0,0)
	\put (5,0) {\line (0,1) {10}}
	\put (5,0) {\circle*{4}}
	\put (5,10) {\circle*{4}}
\end{picture}}\hat{P}_{ij} + \frac{J}{2}'\sum_{
\begin{picture}(11,11)(0,0)
	\put (0,0) {\line (1,1) {10}}
	\put (0,0) {\circle*{4}}
	\put (10,10) {\circle*{4}}
\end{picture}}\hat{P}_{ij}\nonumber\\ 
&+&  \frac{K}{S^2} \sum_{\begin{picture}(11,11)(0,0)
 \put (0,0) {\line (1,0) {10}}
        \put (0,10) {\line (1,0) {10}}
        \put (0,0) {\line (0,1) {10}}
        \put (10,0) {\line (0,1) {10}}
        \put (0,10) {\circle*{4}}
        \put (10,10) {\circle*{4}}
        \put (0,0) {\circle*{4}}
        \put (10,0) {\circle*{4}}
\end{picture}}\bigg (\hat{P}_{ijkl} + \hat{P}_{lkji}\bigg )\nonumber\\ 
&+& \frac{K'}{S^2}\sum_{\begin{picture}(11,11)(0,0)
 \put (-5,0) {\line (1,0) {10}}
        \put (5,10) {\line (1,0) {10}}
        \put (-5,0) {\line (1,1) {10}}
        \put (5,0) {\line (1,1) {10}}
        \put (-5,0) {\circle*{4}}
        \put (15,10) {\circle*{4}}
        \put (5,0) {\circle*{4}}
        \put (5,10) {\circle*{4}}
\end{picture}}\bigg (\hat{P}_{ijkl} + \hat{P}_{lkji}\bigg ) + \frac{K'}{S^2}\sum_{\begin{picture}(11,11)(0,0)
 \put (0,-10) {\line (0,1) {10}}
        \put (0,-10) {\line (1,1) {10}}
        \put (0,0) {\line (1,1) {10}}
        \put (10,0) {\line (0,1) {10}}
        \put (10,10) {\circle*{4}}
        \put (10,0) {\circle*{4}}
        \put (0,0) {\circle*{4}}
        \put (0,-10) {\circle*{4}}
\end{picture}}\bigg (\hat{P}_{ijkl} + \hat{P}_{lkji}\bigg )\nonumber\\
\end{eqnarray}
\noindent where $J$ and $J'$ measure the relative strengths of the nearest neighbour and next-nearest neighbour spin exchange, while $K$, and $K'$ measure the relative strengths of the ring exchange terms on four-sites (since there are three ways to have four-site rings on the anisotropic triangular lattice c.f. Fig. \ref{fig:schematic}). Note that we defined the ring exchange coupling constants $K/S^2$ and $K'/S^2$ to have a meaningful semiclassical limit for $S \to \infty$, i.e., so that the two-spin exchanges are not negligible with respect to the four-spin exchange terms in this limit. The permutation operator which exchanges two spins on sites $i$ and $j$ is given by $\hat{P}_{ij} = 2{\bf S}_{i} \cdot {\bf S}_{j} + \frac{1}{2}$ and $\hat{\bf S}_{i}$ is the usual spin operator on site $i$, while $\hat{P}_{ijkl}=\hat{P}_{ij}\hat{P}_{jk}\hat{P}_{kl}$ cyclically permutes four spins around a plaquette, cf. Fig. \ref{fig:schematic}(b). At this point it is helpful to note that, to lowest order in $t/U$ and $t'/U$, $K'/K = J'/J$. In this work we take this equality to hold, primarily to limit the size of the parameter space of the model.

\begin{figure*}
 \begin{center}
	\includegraphics[width=0.66\columnwidth, clip]{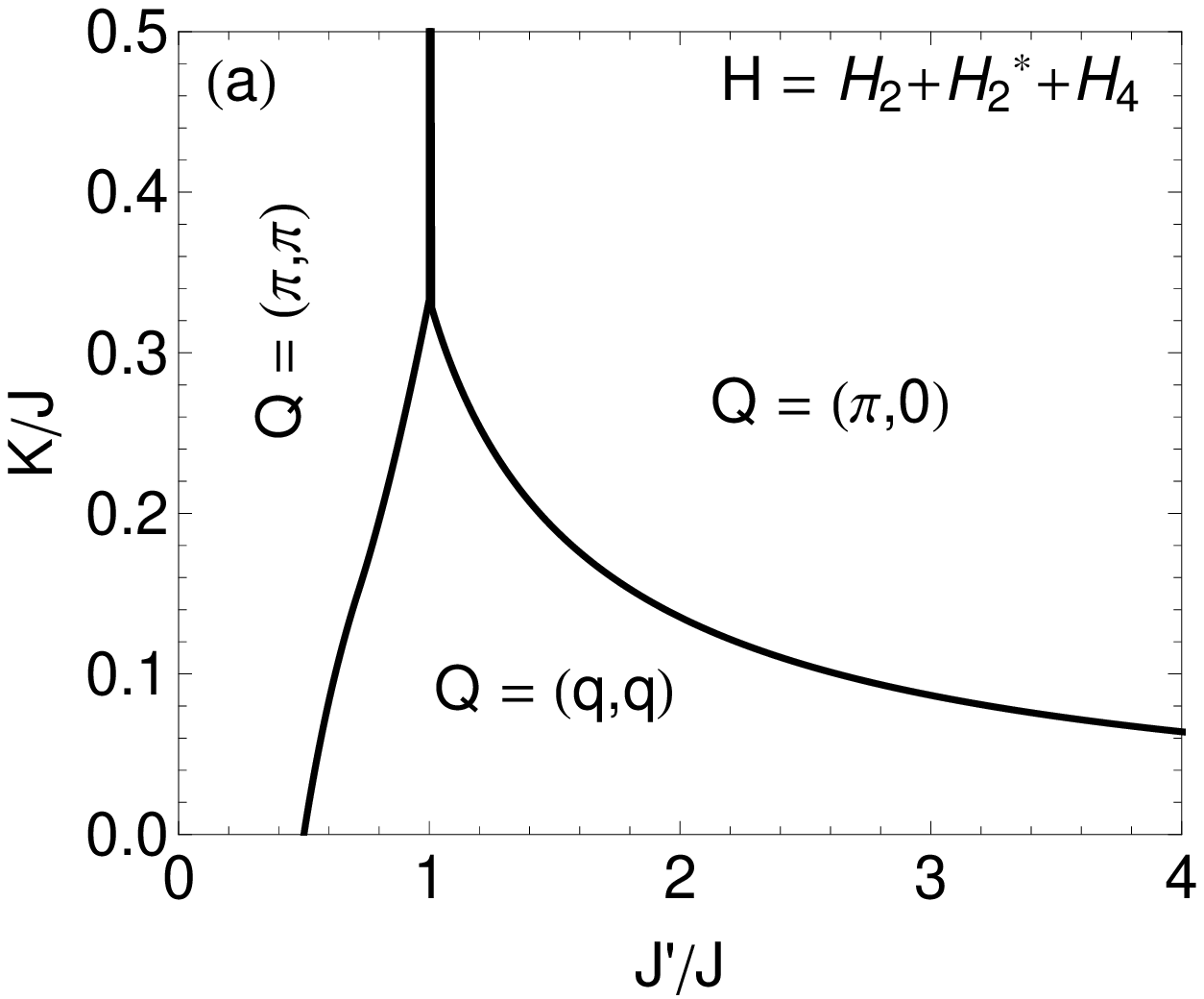}
	\includegraphics[width=0.66\columnwidth, clip]{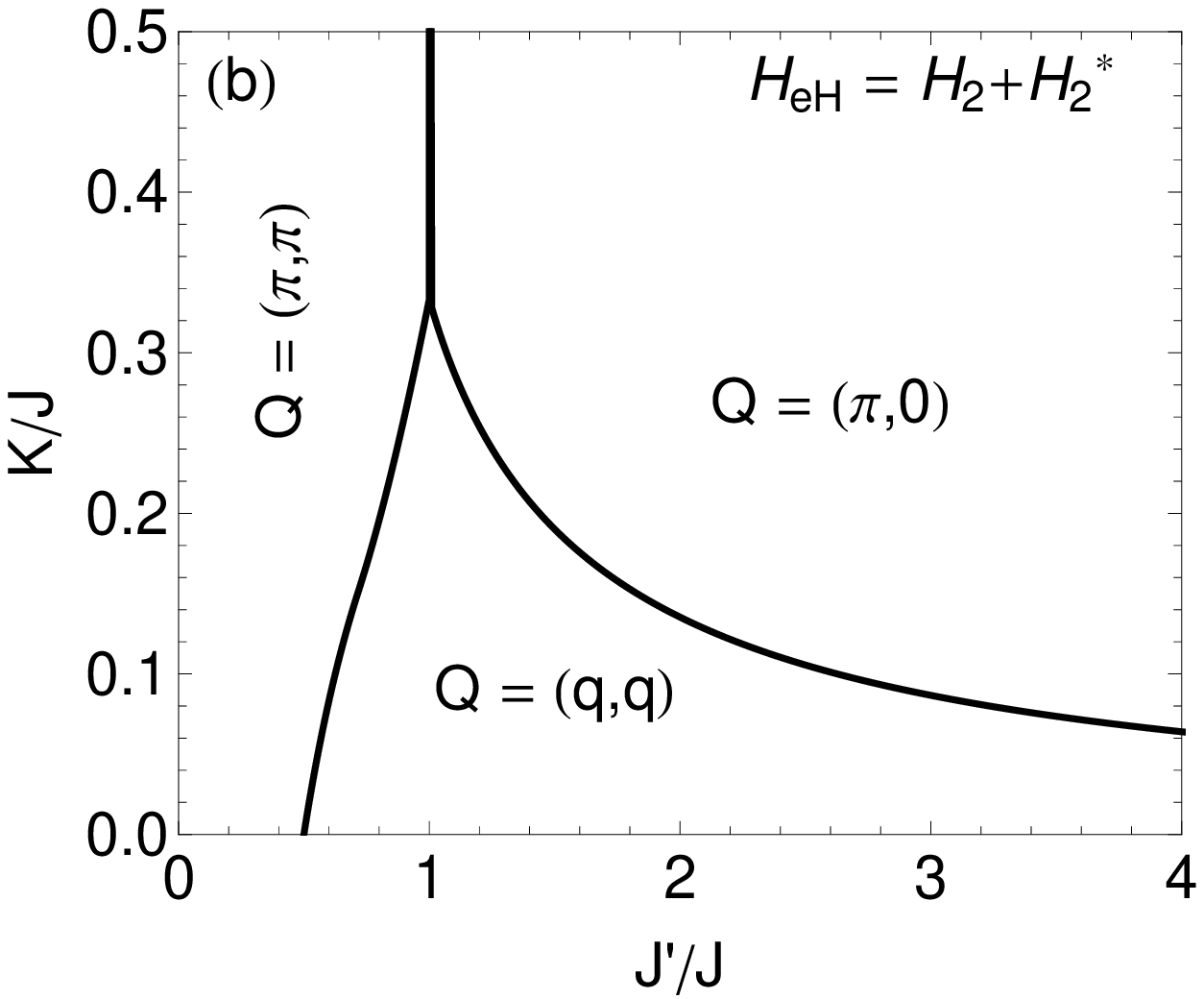}
	\includegraphics[width=0.66\columnwidth, clip]{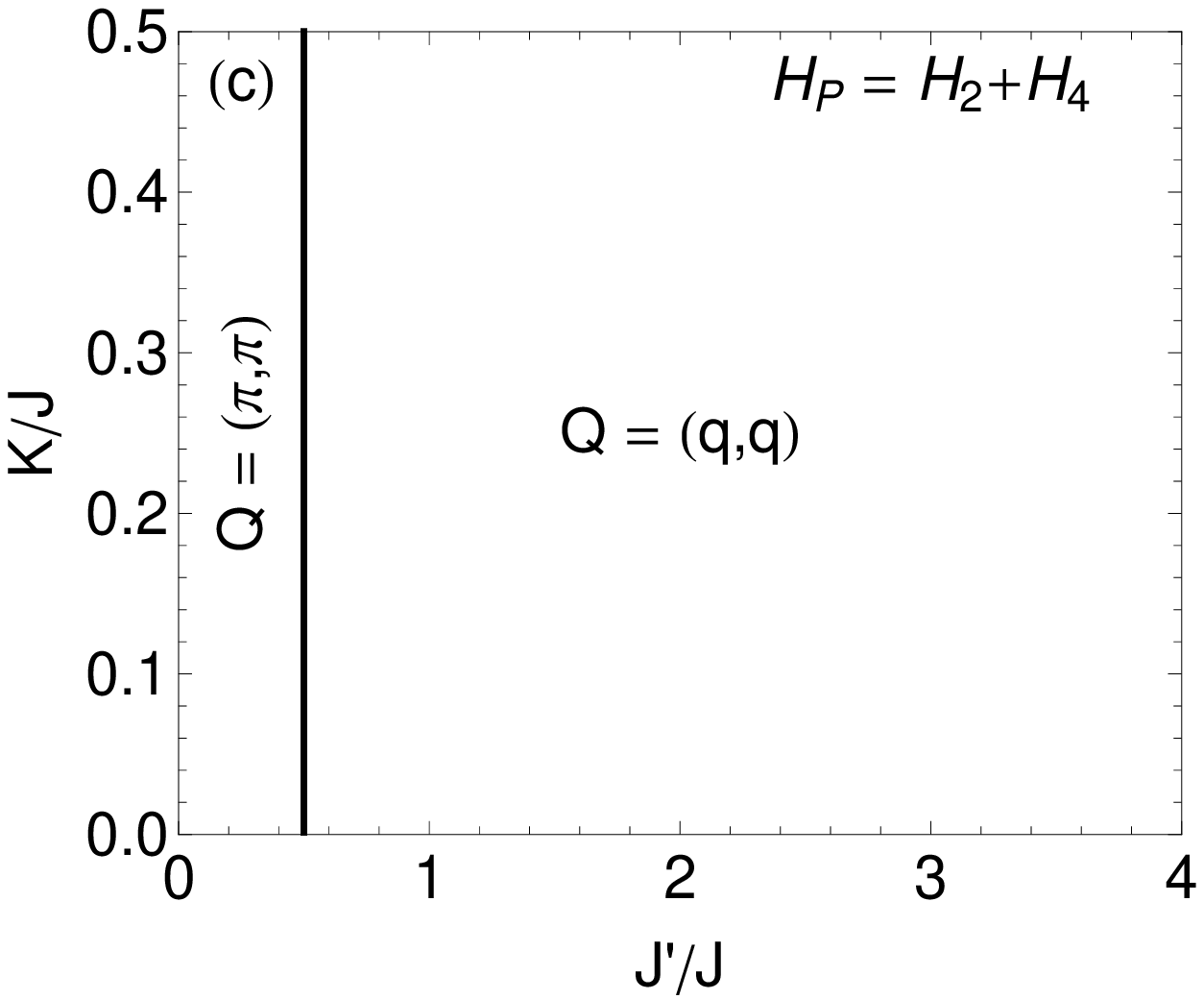}
    \end{center}
    \caption{Classical phase diagrams for the (a) full, (b) extended Heisenberg and (c) plaquette models. The four-spin terms in the Hamiltonian $\hat{H}_{4}$ contribute as trivial constants classically, resulting in the full and extended Heisenberg models having identical classical phase diagrams.}
    \label{fig:phasering}
\end{figure*}

\section{Classical Phase Diagram}

\label{sec:Classical}

Writing out the Hamiltonian in terms of spin operators gives:
\begin{eqnarray}
\label{eq:spinHamiltonianterms}
\hat{H} &=& \hat{H}_{2} + \hat{H}_{2}^{*} + \hat{H}_{4}
\end{eqnarray}
\noindent The first term $\hat{H}_{2}$ arises from the usual Heisenberg exchange terms on the anisotropic triangular lattice
\begin{eqnarray}
\label{eq:HJ}
\hat{H}_{2} &=& J\sum_{
\begin{picture}(17,10)(-2,-2)
	\put (0,0) {\line (1,0) {12}}
	\put (0,0) {\circle*{5}}
	\put (12,0) {\circle*{5}}
\end{picture}}{\bf S}_{i} \cdot {\bf S}_{j}  + J\sum_{
\begin{picture}(17,10)(-2,-2)
	\put (6,-6) {\line (0,1) {12}}
	\put (6,-6) {\circle*{5}}
	\put (6,6) {\circle*{5}}
\end{picture}}{\bf S}_{i} \cdot {\bf S}_{j} + J'\sum_{
\begin{picture}(17,10)(-2,-2)
	\put (2,-6) {\line (1,1) {12}}
	\put (2,-6) {\circle*{5}}
	\put (14,6) {\circle*{5}}
\end{picture}}{\bf S}_{i} \cdot {\bf S}_{j}\nonumber\\
\end{eqnarray}
\noindent The next term $\hat{H}_{2}^{*}$ arises from the two-spin terms from the four-spin permutation operators. Physically it dresses the nearest-neighbour and next-nearest-neighbour exchange strengths and induces next-next nearest neighbour contributions in the Heisenberg model: 
\begin{eqnarray}
\label{eq:HK}
\hat{H}_{2}^{*} &=& (2K + 3K')\sum_{
\begin{picture}(17,10)(-2,-2)
	\put (0,0) {\line (1,0) {12}}
	\put (0,0) {\circle*{5}}
	\put (12,0) {\circle*{5}}
\end{picture}}{\bf S}_{i} \cdot {\bf S}_{j}  + (2K + 3K')\sum_{
\begin{picture}(17,10)(-2,-2)
	\put (6,-6) {\line (0,1) {12}}
	\put (6,-6) {\circle*{5}}
	\put (6,6) {\circle*{5}}
\end{picture}}{\bf S}_{i} \cdot {\bf S}_{j}\nonumber\\ 
&+& (K + 4K')\sum_{
\begin{picture}(17,10)(-2,-2)
	\put (2,-6) {\line (1,1) {12}}
	\put (2,-6) {\circle*{5}}
	\put (14,6) {\circle*{5}}
\end{picture}}{\bf S}_{i} \cdot {\bf S}_{j} + K\sum_{
\begin{picture}(17,10)(-2,-2)
	\put (2,6) {\line (1,-1) {12}}
	\put (2,6) {\circle*{5}}
	\put (14,-6) {\circle*{5}}
\end{picture}}{\bf S}_{i} \cdot {\bf S}_{j}\nonumber\\ 
&+& K'\sum_{
\begin{picture}(17,10)(-2,-2)
	\put (-2,-4) {\line (2,1) {18}}
	\put (-2,-4) {\circle*{5}}
	\put (16,6) {\circle*{5}}
\end{picture}}{\bf S}_{i} \cdot {\bf S}_{j} + K'\sum_{
\begin{picture}(17,10)(-2,-2)
	\put (2,-12) {\line (1,2) {10}}
	\put (2,-12) {\circle*{5}}
	\put (10,6) {\circle*{5}}
\end{picture}}{\bf S}_{i} \cdot {\bf S}_{j}\nonumber\\
\end{eqnarray}
\noindent The final term consists of the four-spin plaquette terms on the anisotropic triangular lattice.
\begin{eqnarray}
\label{eq:HKprime}
\hat{H}_{4} &=& \frac{K}{S^2}\sum_{\begin{picture}(26,15)(-2,-2)
 \put (6,0) {\line (1,0) {12}}
        \put (6,10) {\line (1,0) {12}}
        \put (6,0) {\line (0,1) {12}}
        \put (18,0) {\line (0,1) {12}}
        \put (6,10) {\circle*{5}}
        \put (18,10) {\circle*{5}}
        \put (6,0) {\circle*{5}}
        \put (18,0) {\circle*{5}}
\end{picture}} \hat{T}_{ijkl} + \frac{K'}{S^2}\sum_{\begin{picture}(26,15)(-2,-2)
 \put (-6,0) {\line (1,0) {18}}
        \put (4,12) {\line (1,0) {18}}
        \put (-6,0) {\line (1,1) {12}}
        \put (10,0) {\line (1,1) {12}}
        \put (6,12) {\circle*{5}}
        \put (22,12) {\circle*{5}}
        \put (-6,0) {\circle*{5}}
        \put (10,0) {\circle*{5}}
\end{picture}} \hat{T}_{ijkl} + \frac{K'}{S^2} \sum_{\begin{picture}(26,15)(-2,-2)
 \put (6,-12) {\line (0,1) {12}}
        \put (6,-12) {\line (1,1) {12}}
        \put (6,0) {\line (1,1) {12}}
        \put (18,-2) {\line (0,1) {12}}
        \put (6,-12) {\circle*{5}}
        \put (18,12) {\circle*{5}}
        \put (6,0) {\circle*{5}}
        \put (18,0) {\circle*{5}}
\end{picture}}\hat{T}_{ijkl}\nonumber\\
\end{eqnarray}
\noindent where
\begin{eqnarray}
\label{eq:Tijlm}
\hat{T}_{ijkl} &=& ({\bf S}_{i} \cdot {\bf S}_{j})({\bf S}_{l} \cdot {\bf S}_{m}) + ({\bf S}_{i} \cdot {\bf S}_{m})({\bf S}_{j} \cdot {\bf S}_{l})\nonumber\\
&-& ({\bf S}_{i} \cdot {\bf S}_{l})({\bf S}_{j} \cdot {\bf S}_{m})
\end{eqnarray}
We are interested in the relative effects of each of the terms in Eq. \eqref{eq:spinHamiltonianterms}. We will therefore explicitly compare three models: (i) the full model given by  Eq. \eqref{eq:spinHamiltonianterms}, (ii) the extended Heisenberg model, $\hat{H}_{eH} = \hat{H}_{2} + \hat{H}_{2}^{*}$, and finally (ii) the four-spin plaquette model $\hat{H}_{P} = \hat{H}_{2} + \hat{H}_{4}$. From now on, these three models will be referred to as the ``full'', ``extended Heisenberg", and ``plaquette'' models. 

The four-spin terms in  Eq. \eqref{eq:Tijlm} are decoupled in a leading order mean-field approximation taking
\begin{eqnarray}
\label{eq:twospinexpectation}
\langle {\bf S}_{\alpha} \cdot {\bf S}_{\beta}\rangle &=&  S^2 \cos({\bf Q} \cdot {\bm \delta_{\alpha \beta}})
\end{eqnarray}
\noindent so that 
\begin{widetext}
\begin{eqnarray}
\label{eq:quarticspinsHartreeFockshorthand}
\frac{1}{S^2}({\bf S}_{\alpha} \cdot {\bf S}_{\beta})({\bf S}_{\gamma} \cdot {\bf S}_{\delta}) &=& \cos({\bf Q} \cdot {\bm \delta_{\alpha \beta}}){\bf S}_{\gamma} \cdot {\bf S}_{\delta}
+ \cos({\bf Q} \cdot {\bm \delta_{\gamma \delta}}){\bf S}_{\alpha} \cdot {\bf S}_{\beta}
- S^2\cos({\bf Q} \cdot {\bm \delta_{\alpha \beta}})\cos({\bf Q} \cdot {\bm \delta_{\gamma \delta}}) 
\end{eqnarray}

At this level of approximation ring-exchange contributes by dressing the effective two spin exchange and, in particular, introduces additional long-range frustrated interactions
\begin{eqnarray}
\label{eq:dressedexchange}
J_{\hat{\bf x}} &=& J + 2K + 3K' + 2(K+K')\cos(Q_{x})
 - K'\cos(Q_{x}+2Q_{y})\nonumber\\
J_{\hat{\bf y}} &=& J + 2K + 3K' + 2(K+K')\cos(Q_{y})
 - K'\cos(2Q_{x}+Q_{y})\nonumber\\
J_{\hat{\bf x} + \hat{\bf y}} &=& J' + K + 4K' + 4K'\cos(Q_{x}+Q_{y})
 - K\cos(Q_{x}-Q_{y})\nonumber\\
J_{\hat{\bf x} - \hat{\bf y}} &=& K(1 - \cos(Q_{x}+Q_{y})) \nonumber\\
J_{2\hat{\bf x} + \hat{\bf y}} &=& K'(1 - \cos(Q_{y})) \nonumber\\
J_{\hat{\bf x} + 2\hat{\bf y}} &=& K'(1 - \cos(Q_{x}))
\end{eqnarray}
\noindent where $J_{\hat{\bm \eta}}$ describes an antiferromagnetic exchange interaction in the $ \hat{\bm \eta}$ direction. Here $\hat{\bf x}$ and $\hat{\bf y}$ are vectors of length one lattice spacing in the $x$ and $y$ directions respectively (cf. Fig. \ref{fig:schematic}(a)). All other $J_{\hat{\bm\eta}}$ are zero. 

The classical ground-state energy per site for the full model is
\begin{eqnarray}
\label{eq:EGS0}
\frac{E_{GS}^{(0)}}{NS^2} &=& J\bigg (\cos(Q_{x}) + \cos(Q_{y})\bigg ) +J'\cos(Q_{x} + Q_{y})
+ K\bigg (1 + 2\cos(Q_{x}) + 2\cos(Q_{y}) + 2\cos(Q_{x})\cos(Q_{y})\bigg )\nonumber\\ 
&&+ K'\bigg (2 + 3\cos(Q_{x}) + 3\cos(Q_{y}) + 4\cos(Q_{x} + Q_{y})
+ \cos(2Q_{x} + Q_{y}) + \cos(Q_{x} + 2Q_{y}) \bigg ).
\end{eqnarray}
\noindent We note that the four-spin contibution  (Eq. \eqref{eq:HKprime}) to the classical ground state energy leads to a trivial constant $K + K'$ and the energies (and thence all other properties) of the full and extended Heisenberg models are identical. 

To calculate the classical phase diagram we considered N\'{e}el, collinear, and spiral phases with ordering vectors  ${\bf Q} = (\pi, \pi)$, ${\bf Q} = (\pi, 0)$, and  ${\bf Q} = (q, q)$ respectively. We also considered several other phases including the diagonal, incommensurate spiral, columnar dimer and diagonal dimer phases with ${\bf Q} = (\pi/2, \pi/2)$, ${\bf Q} = (\pi, q)$, ${\bf Q} = (q-\pi, q)$ and ${\bf Q} = (\pi-2q, q)$ respectively when considering the classical phase diagram but these are all higher in energy. The ordering vector for the commensurate spiral phase $q$ is found by minimizing  Eq. \eqref{eq:EGS0} with respect to $q$ assuming $Q_{x} = Q_{y} = q$. We find that the spiral ordering vector for the full and extended Heisenberg models depend on ring exchange, and are given by

\begin{eqnarray}
\label{eq:QSpiral}
q = \cos^{-1}\bigg ( -\frac{J' + K + 4K' - \sqrt{(J' + K + 4K')^2 - 12(J + 2K)K'}}{12K'} \bigg )
\end{eqnarray}
\end{widetext}
\noindent whereas for the plaquette model we find 
\begin{eqnarray}
\label{eq:QSpiral4spin}
q = \cos^{-1}\bigg ( -\frac{J}{2J'} \bigg )
\end{eqnarray}
In Fig. \ref{fig:phasering} we plot the classical phase diagrams obtained for the three models. For $K = 0$ we observe a transition from N\'{e}el to spiral order with $q = \cos^{-1}(-J/2J')$ for $J'/J = 0.5$, consistent with previous studies of the Heisenberg model \cite{Merino99, Trumper99}. With increasing ring exchange the N\'{e}el and collinear phases are stabilized, while the spiral order is destabilized in the full and extended Heisenberg models. Even at the classical level the spiral phase is most stable to ring exchange when $J' = J$.  We will see below that this stabilisation of the spiral phase is reflected in the quantum calculations.

For the plaquette model, we find that the critical point between the N\'{e}el and spiral phases is independent of ring exchange classically since the ordering vector for the spiral phase in this model  (Eq. \eqref{eq:QSpiral4spin}) is independent of ring exchange. The collinear phase is not observed in the plaquette models since it is always higher in energy than the spiral ordering.

\section{Linear Spin-Wave Theory}

\label{sec:LSWT}

We study the quantum phase diagram and elementary excitations for $S = 1/2$ at $T = 0$ using linear spin-wave theory. It is convenient \cite{Miyake92, Singh92} to assume that the spins lie in the $x-z$ plane and rotate the quantum projection axis of the spins at each site along its classical direction:
\begin{eqnarray}
\label{eq:rotatespins}
\hat{S}_{i}^{x'} &=& \tilde{S}_{i}^{x}\cos(\theta_{i}) + \tilde{S}_{i}^{z}\sin(\theta_{i})\nonumber\\
\hat{S}_{i}^{y'} &=& \tilde{S}_{i}^{y}\nonumber\\
\hat{S}_{i}^{z'} &=& -\tilde{S}_{i}^{x}\sin(\theta_{i}) + \tilde{S}_{i}^{z}\cos(\theta_{i})
\end{eqnarray}
\noindent Here $\theta_{i} = {\bf Q} \cdot {\bf r}_{i}$, where ${\bf r}_{i}$ is the position of the $i^\textrm{th}$ spin and $ {\bf Q} = (Q_{x}, Q_{y})$ is the ordering vector of the lattice. This simplifies the spin-wave treatment with the result that only one, rather than three, species of boson is required to describe the spin operators.

The bosonization of the spin operators is performed via the Holstein-Primakov transformation
\begin{eqnarray}
\label{eq:HolsteinPrimakov}
\tilde{S}_{i}^{z} &=& S - \hat{a}_{i}^{\dagger} \hat{a}_{i}\nonumber\\
\tilde{S}_{i}^{+} &=& \sqrt{2S - \hat{a}_{i}^{\dagger} \hat{a}_{i}}\hat{a}_{i}\nonumber\\
\tilde{S}_{i}^{-} &=& \hat{a}_{i}^{\dagger}\sqrt{2S - \hat{a}_{i}^{\dagger} \hat{a}_{i}}
\end{eqnarray}
\noindent where $\tilde{S}_{i}^{\pm} = \tilde{S}_{i}^{x} \pm i \tilde{S}_{i}^{y}$ with subsequent expansion of square roots in powers of  $a_{i}^{\dagger} a_{i}/(2S)$. Linear spin-wave theory takes the leading order terms in a $1/S$ expansion, which describe noninteracting spin waves.  Performing a Fourier transform of the bosonic operators results in the following Hamiltonian
\begin{equation}
\label{eq:quadraticHamiltonian}
\hat{H}_{LSWT} = 2S\sum_{\bf k} \bigg [A_{\bf k} a_{\bf k}^{\dagger} a_{\bf k} - \frac{B_{\bf k}}{2}(a_{\bf k}^{\dagger} a_{-\bf k}^{\dagger} + a_{\bf k} a_{-\bf k}) \bigg ]
\end{equation}
\noindent where
\begin{eqnarray}
\label{eq:modifiedcoeff}
A_{\bf k} &=& \sum_{\eta} \frac{J_{\eta}}{2} \cos ({\bf k} \cdot {\bm \delta}_{\eta})(\cos({\bf Q} \cdot {\bm \delta}_{\eta}) + 1) - J_{\eta} \cos ({\bf Q} \cdot {\bm \delta}_{\eta})\nonumber\\
B_{\bf k} &=& \sum_{\eta} \frac{J_{\eta}}{2}  \cos({\bf k} \cdot {\bm \delta}_{\eta})(1 - \cos ({\bf Q} \cdot {\bm \delta}_{\eta}))
\end{eqnarray}

\begin{figure*}
 \begin{center}
     \includegraphics[width=0.66\columnwidth, clip]{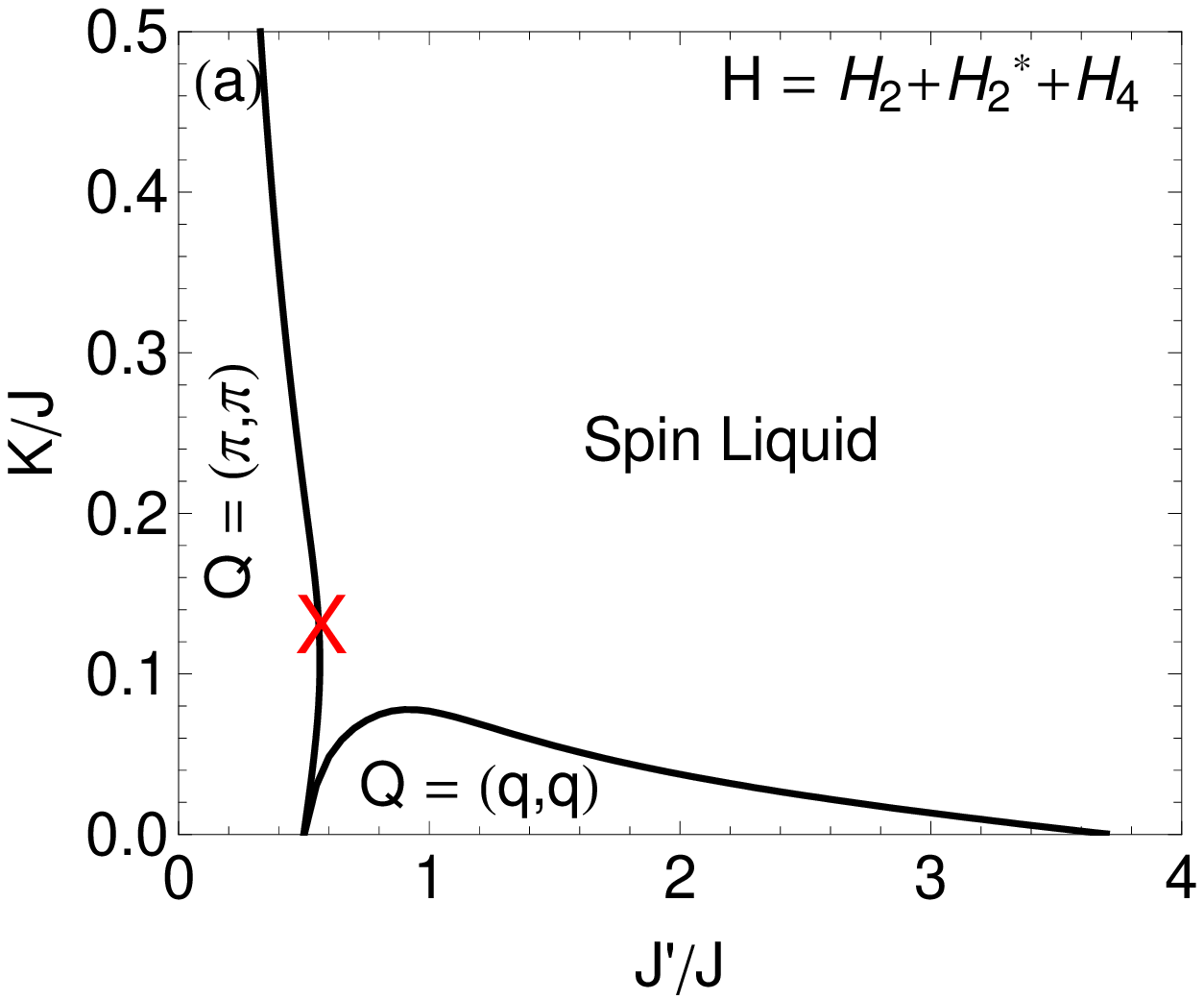}
\includegraphics[width=0.66\columnwidth, clip]{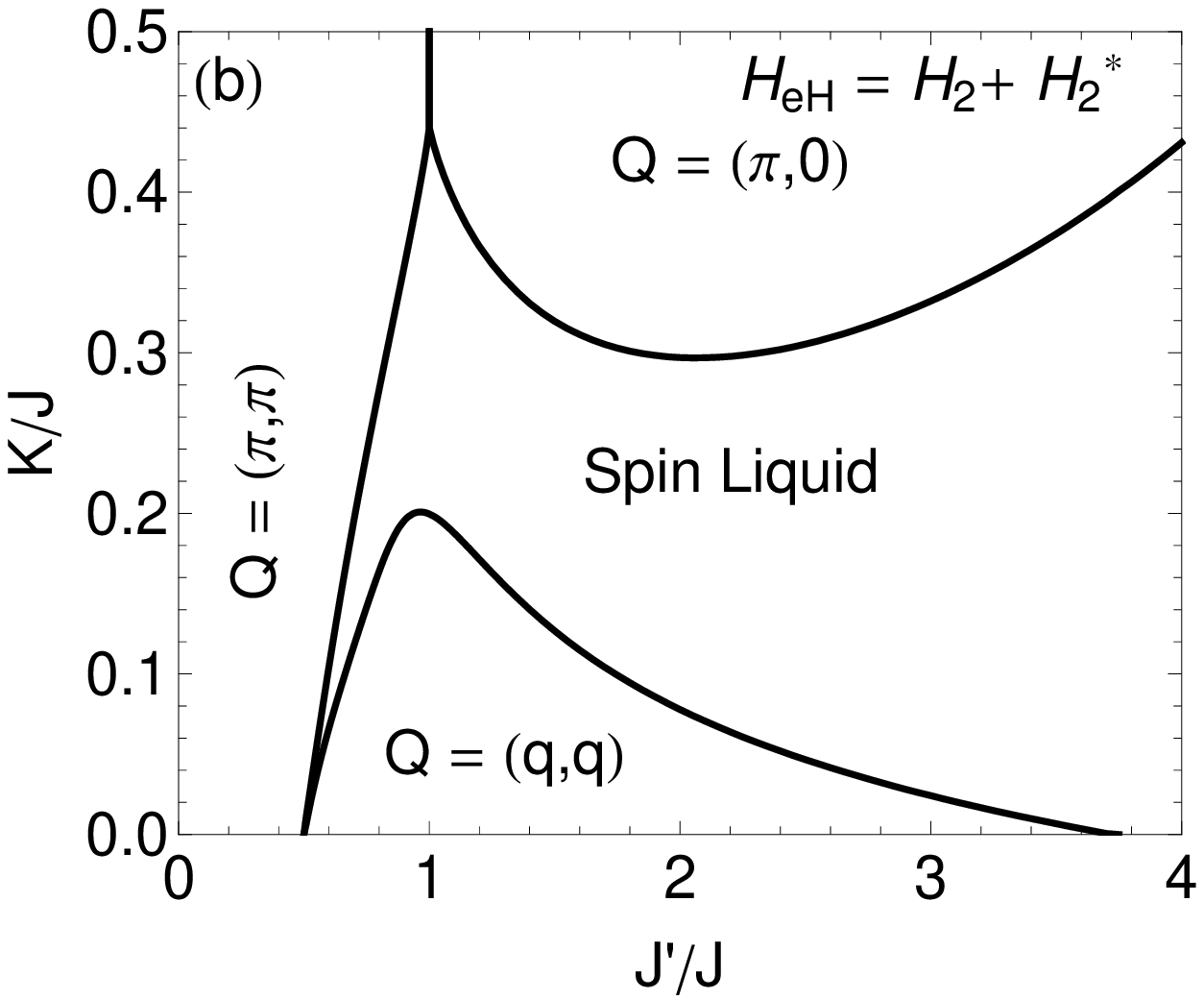}
\includegraphics[width=0.66\columnwidth, clip]{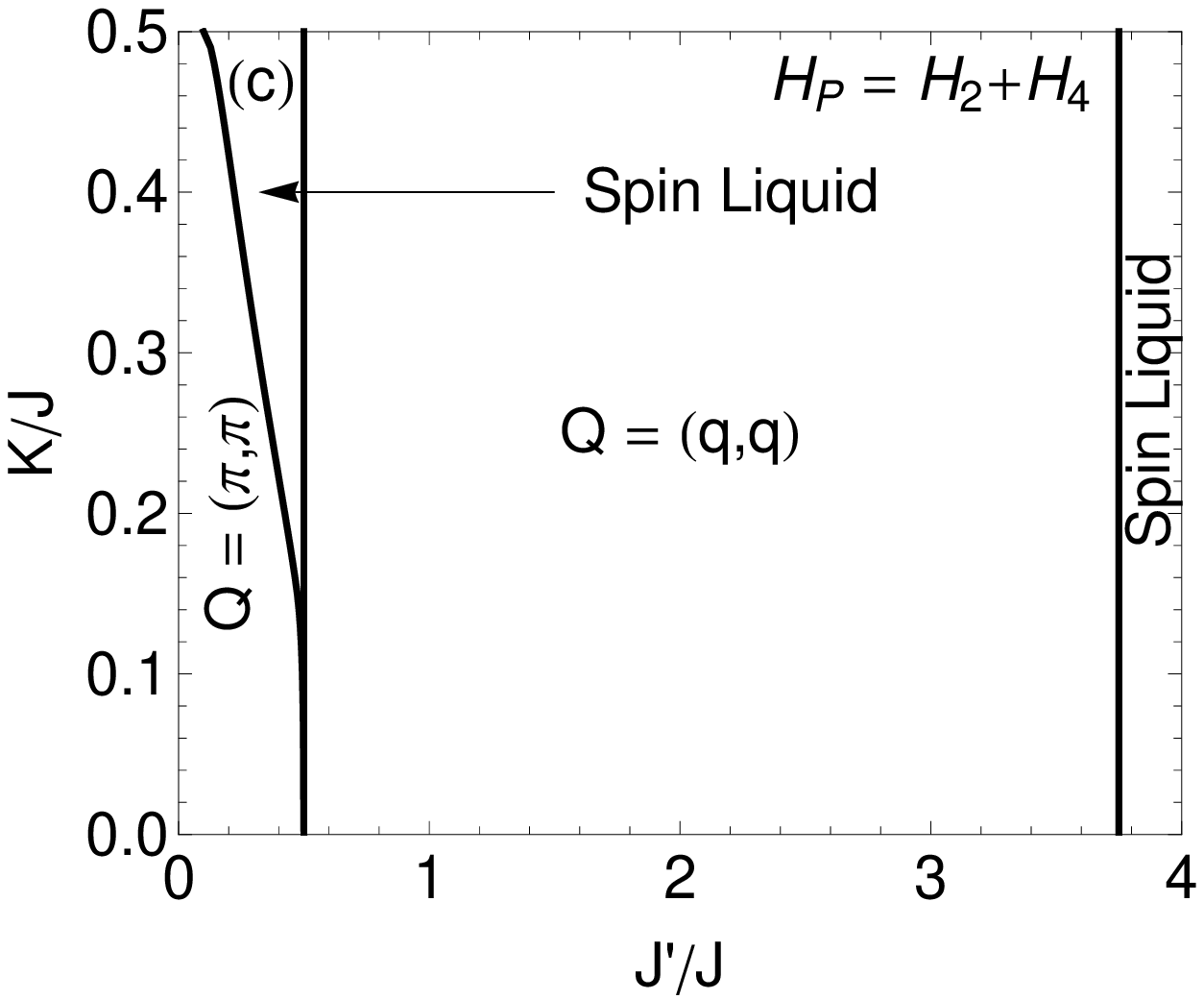}
    \end{center}
    \caption{(Color online) Quantum phase diagrams for (a) the full and extended Heisenberg models and (b) for the plaquette model. It is clear that, even in these semiclassical calculations, quantum fluctuations strongly suppress long-range order in the full and extended Heisenberg models when ring exchange is introduced. In addition we have marked the critical end point where the staggered magnetization transitions from a second order phase transition to a first order phase transition with a red cross.}
    \label{fig:phaseringquantum}
\end{figure*}

\noindent Introducing the Fourier transform of the exchange interaction
\begin{equation}
\label{eq:Jk}
J_{\bf k} = \sum_{\eta} J_{\eta}\cos({\bf k} \cdot \bm\delta_{\eta}) 
\end{equation}
\noindent allows us to conveniently express the functions $A_{\bf k}$ and $B_{\bf k}$ through $J_{\bf k}$ as
\begin{eqnarray}
\label{eq:AkqBkq}
A_{\bf k} &=& \frac{1}{4}\bigg [J_{{\bf Q}+{\bf k}} + J_{{\bf Q}-{\bf k}}\bigg ] + \frac{J_{\bf k}}{2} - J_{\bf Q}\nonumber\\
B_{\bf k} &=&  \frac{J_{\bf k}}{2} - \frac{1}{4}\bigg [J_{{\bf Q}+{\bf k}} + J_{{\bf Q}-{\bf k}}\bigg ]
\end{eqnarray}
\noindent We proceed by diagonalizing  Eq. \eqref{eq:quadraticHamiltonian} via a Bogoliubov transformation
\begin{eqnarray}
\label{eq:Bogoliubov}
a_{\bf k}^{\dagger} &=& u_{\bf k} \alpha_{\bf k}^{\dagger} +  v_{\bf k} \alpha_{-\bf k}\nonumber\\
a_{\bf k} &=& u_{\bf k} \alpha_{\bf k} +  v_{\bf k} \alpha_{-\bf k}^{\dagger}
\end{eqnarray}
\noindent where
\begin{eqnarray}
\label{eq:Bogoliubovcoeff}
u_{\bf k} &=& \sqrt{\frac{A_{\bf k}}{\omega_{\bf k}} + \frac{1}{2}}\nonumber\\
v_{\bf k} &=& sgn(B_{\bf k})\sqrt{\frac{A_{\bf k}}{\omega_{\bf k}} - \frac{1}{2}}
\end{eqnarray}
\noindent which yields
\begin{equation}
\label{eq:diagonalizedHamiltonian}
\hat{H} = E_{GS}^{(0)} + \frac{1}{2}\sum_{\bf k} (\omega_{\bf k} - A_{\bf k}) + \sum_{\bf k} \omega_{\bf k} \hat{\alpha}_{\bf k}^{\dagger} \hat{\alpha}_{\bf k}
\end{equation}
\noindent where the spin-excitation spectrum, $\omega_{\bf k}$, reads
\begin{eqnarray}
\label{eq:spinspectra}
\omega_{\bf k} = 2S\sqrt{(J_{\bf k} - J_{\bf Q})([J_{\bf Q + k} + J_{\bf Q-k}]/2 - J_{\bf Q})}
\end{eqnarray}
\noindent It is already clear from Eq. \eqref{eq:spinspectra} that the magnon spectrum has zeros at ${\bf k} = \bm 0$ and ${\bf k} = \pm {\bf Q}$ in the two-dimensional Brillouin zone.

In studying the quantum phase diagram a key quantity is the staggered magnetization,
\begin{eqnarray}
\label{eq:mstag}
m_{s} =\langle \tilde{S}_{i}^{z} \rangle = S + \frac{1}{2} - \frac{1}{8\pi^2}\int_{BZ} \frac{A_{\bf k}}{\omega_{\bf k}}d^2k
\end{eqnarray}
\noindent as  $m_{s}$ vanishes in a a quantum disordered state.

\section{Ground-State Properties}

\label{sec:GroundState}

\subsection{Quantum Phase Diagrams}

The quantum phase diagrams for the three models are shown in Fig. \ref{fig:phaseringquantum}. All three models are equivalent for $K/J = K'/J = 0.0$. This model has been studied via LSWT previously.\cite{Merino99, Trumper99} As in these studies we find a quantum critical point driven by the vanishing of the spin-wave velocity at $J'/J = 0.5$ and a quantum disordered phase for $J'/J \gtrsim 3.75$.

\subsubsection{Full Model}

\begin{figure*}
 \begin{center}
      \includegraphics[width=0.67\columnwidth, trim = 0mm 0mm 0mm 0mm, clip]{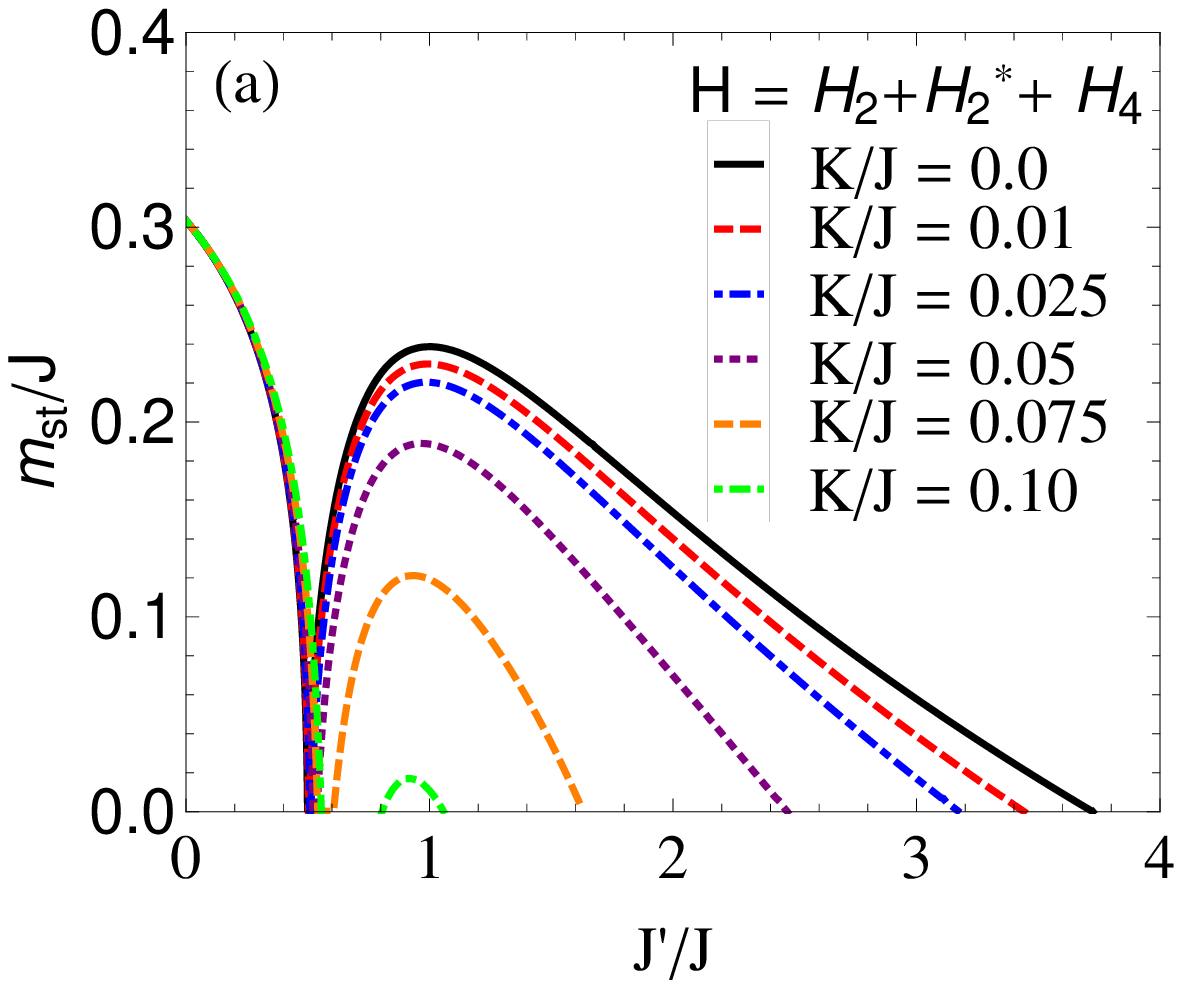}
	\includegraphics[width=0.67\columnwidth, trim = 0mm 0mm 0mm 0mm, clip]{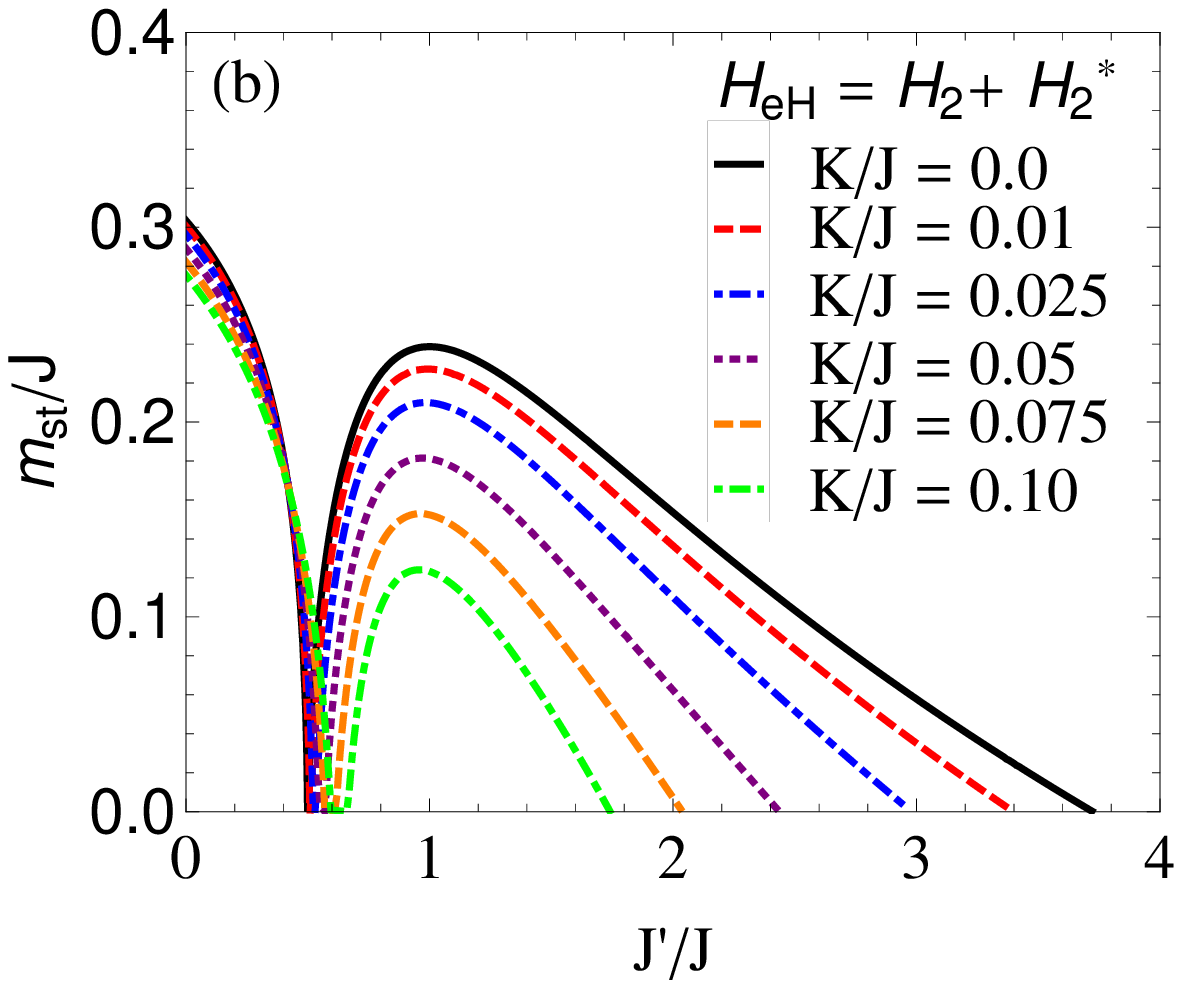}
	\includegraphics[width=0.67\columnwidth, trim = 0mm 0mm 0mm 0mm, clip]{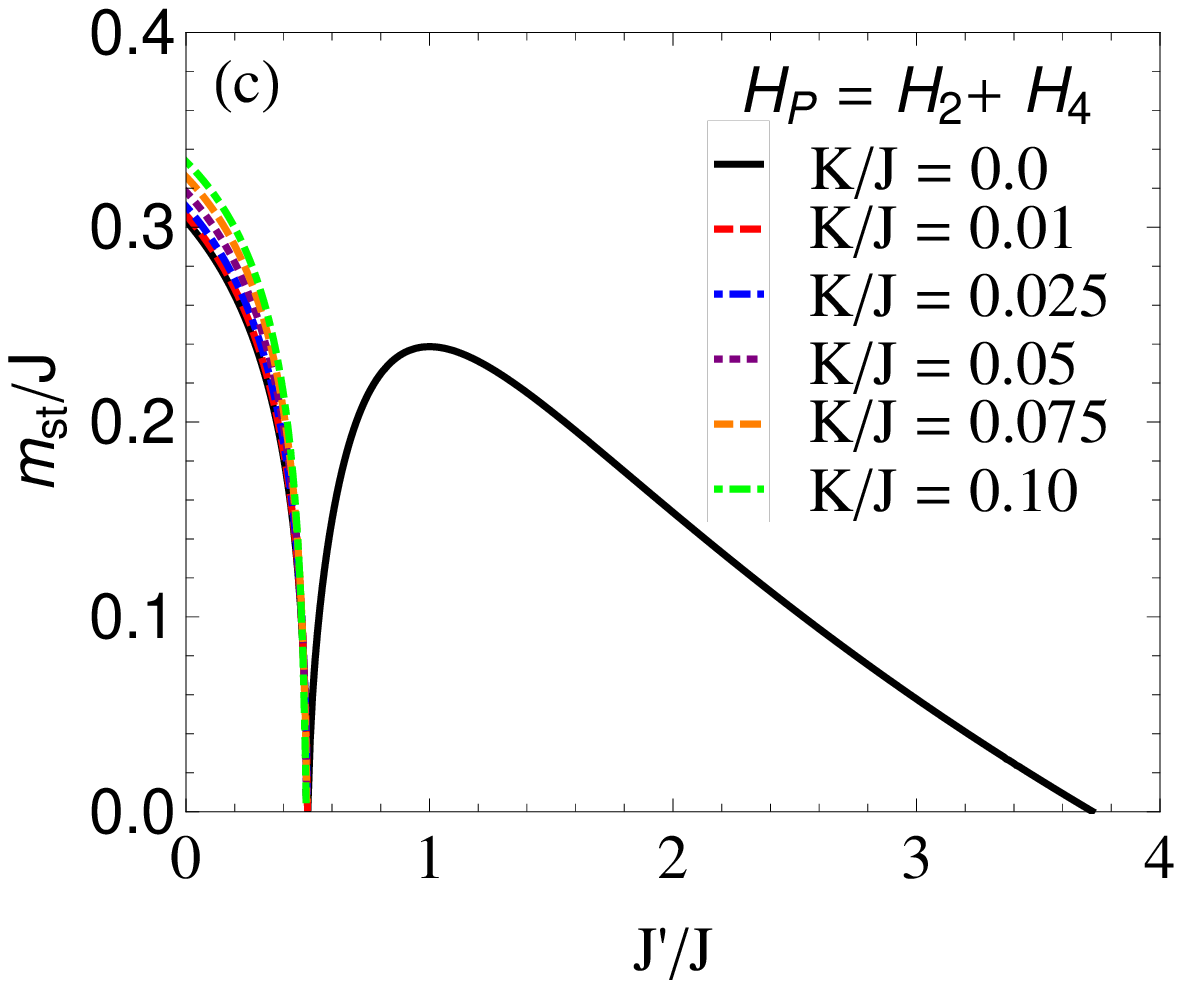}
    \end{center}
    \caption{(Color online) Staggered magnetization for the anisotropic triangular lattice with ring exchange calculated using linear spin-wave theory (LSWT). In each plot we show the N\'{e}el-spiral transition for the (a) full model, (b) extended Heisenberg and (c) plaquette models.}
    \label{fig:mstagmodels}
\end{figure*}

\begin{figure*}
 \begin{center}
      \includegraphics[width=0.67\columnwidth, trim = 0mm 0mm 0mm 0mm, clip]{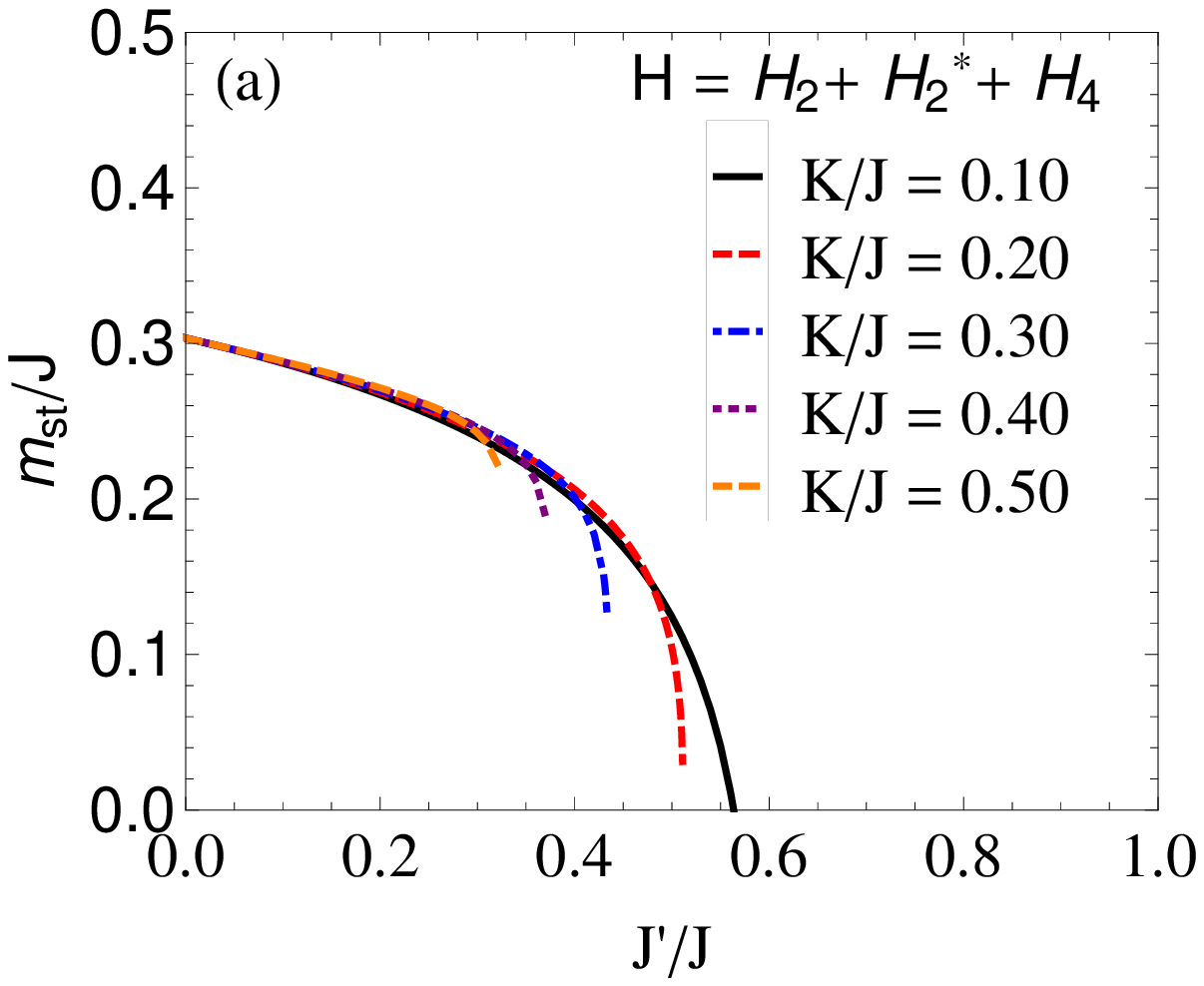}
	\includegraphics[width=0.67\columnwidth, trim = 0mm 0mm 0mm 0mm, clip]{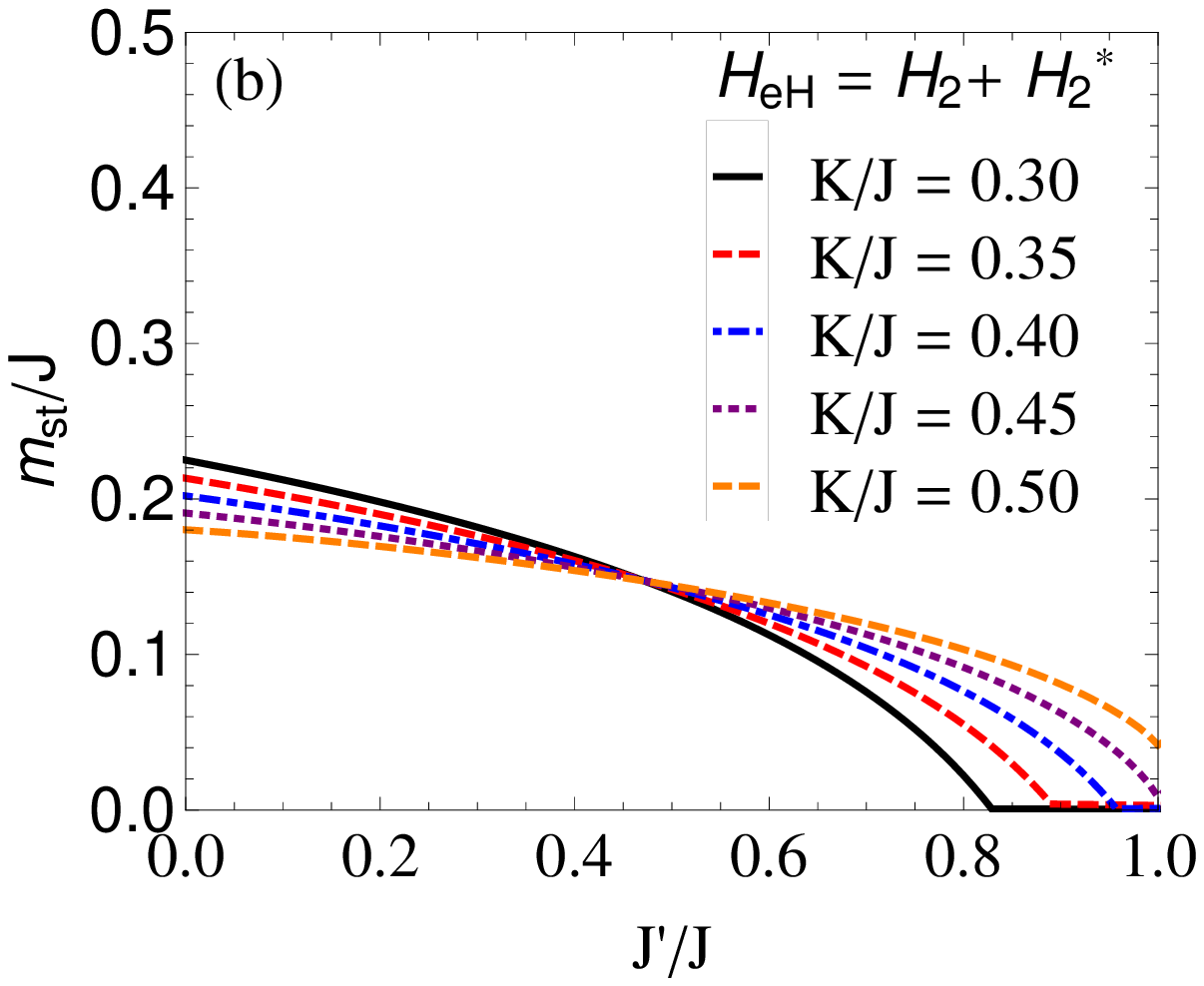}
	\includegraphics[width=0.67\columnwidth, trim = 0mm 0mm 0mm 0mm, clip]{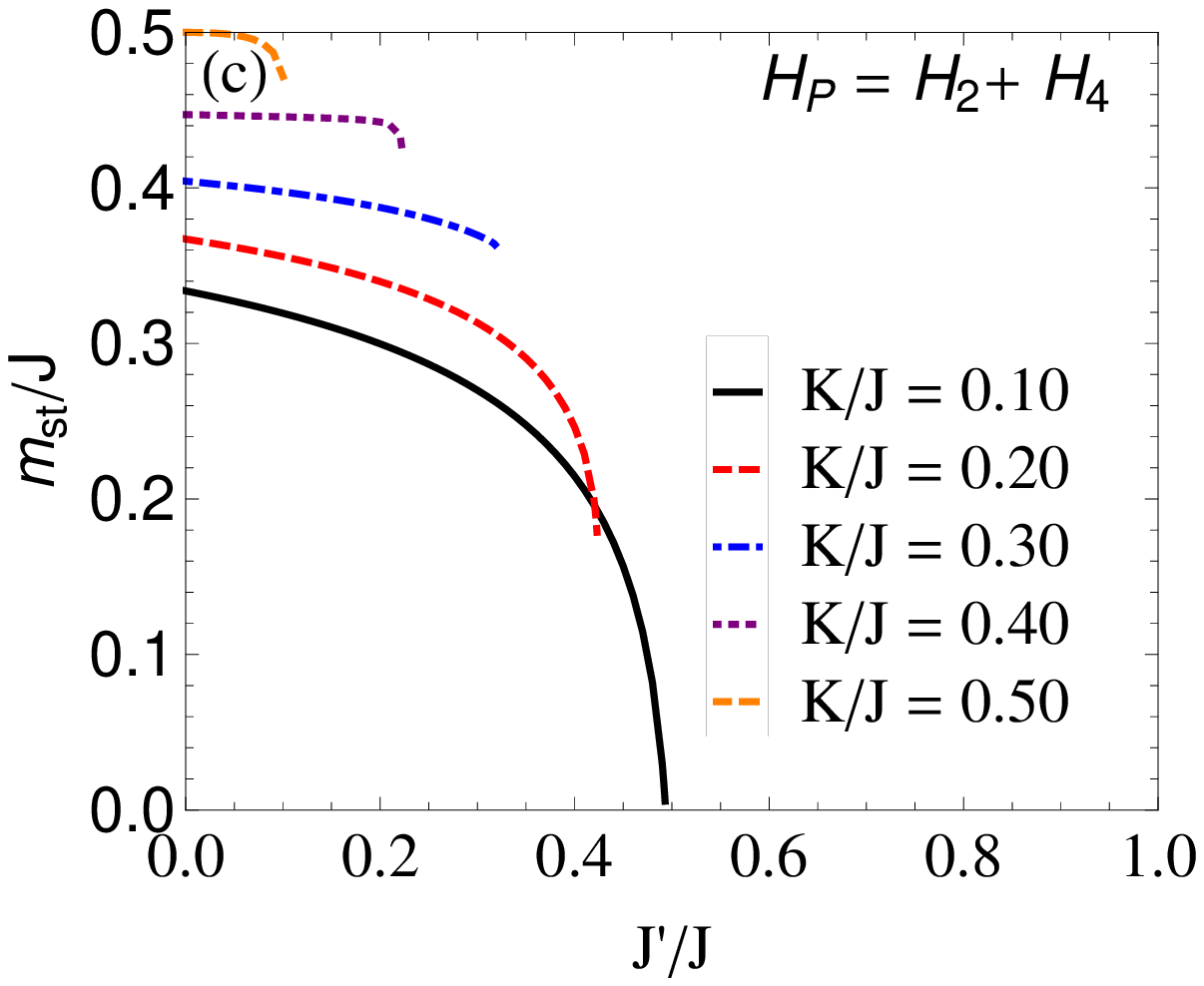}
    \end{center}
    \caption{(Color online) Staggered magnetization for the anisotropic triangular lattice with ring exchange calculated using linear spin-wave theory (LSWT). In each plot we show the destabilization of N\'{e}el order for the (a) full model, (b) extended Heisenberg and (c) plaquette models.}
    \label{fig:vanishmstagmodels}
\end{figure*}

In the full model   (Fig. \ref{fig:phaseringquantum}(a)) ring-exchange initially favours N\'{e}el order until the critical value of ring-exchange, $K_c$, further increase of the ring-exchange coupling destabilizes N\'{e}el order. Similar behaviour was observed by Majumdar {\it et al}. \cite{Majumdar12}. Interestingly for $K<K_c$ the transition between the N\'{e}el and spin liquid phases is second order, but for $K>K_c$ this transition is first order (cf. Figs. \ref{fig:mstagmodels} and \ref{fig:vanishmstagmodels}).

We also found that the spiral phase was dramatically suppressed in the quantum calculations compared to that found classically; the spiral phase only survives up to $K/J = 0.10$ for $J'/J = 1$ in constast to $K/J = 1/3$, found classically. %In the weakly-coupled chain limit of our model $J'/J \gg 1$ we found spiral order persists in the presence of weak frustration from ring exchange ($K'/K \ll 1$). 
In a large region of the quantum phase diagram, the classically stable collinear phase is wiped out and replaced with a spin-liquid phase. 

The behavior that we observe in the quantum phase diagram is highly analogous to the `order-by-disorder' mechanism due to quantum or thermal fluctuations \cite{Villain, Chandra90, Chubukov92,Sheng92}. In this mechanism, the fluctuations determine which ground-states are stabilized and hence selected. %The order-by-disorder mechanism has previously been studied on the anisotropic triangular lattice by Sheng and Henley \cite{Sheng92}. %In our model we find that quantum fluctuations lead to a direct competition between the classically ordered states. In addition to this we find that there is a separation between the N\'{e}el and spiral phases for $K/J > 0$. This is most notably seen for the $K/J = 0.05$ and $K/J = 0.075$ plots in Fig. \ref{fig:mstagmodels}(a).

For the isotropic case ($J'/J = 1$) we found spiral order persists for $K/J < 0.1$. Previously Motrunich \cite{Motrunich05} predicted spiral order to be preserved for small $K/J \lesssim 0.14-0.20$ \cite{fn1} but be destroyed for larger $K/J$ leading to a gapped spin-liquid for $K/J > 0.28$. Additionally, Kubo and Momoi considered numerous mean-field ground states with up to 144 sublattices \cite{Kubo03}. In zero magnetic field they found the spiral phase with $120^{\circ}$ order exists for $K/J < 0.1$. Our results  are consistent with both of these studies. 

The most striking feature of the phase diagram is that, even in this semi-classical theory, quantum fluctuations destroy long range magnetic order over large areas of the phase diagram. These quantum disordered regions occur in the parameter region consistent with DMRG calculations on four-leg triangular ladders.\cite{Block11}  

\subsubsection{Extended Heisenberg and Plaquette Models}

For the extended Heisenberg model  (Fig. \ref{fig:phaseringquantum}(b)) we  find that N\'{e}el order is stabilized by ring exchange and extends to the isotropic case ($J'/J = 1$) for $K/J \approx 0.44$ where it undergoes a first-order transition to the collinear phase. Additionally, we find that the spiral phase was suppressed in the quantum calculations relative to that found classically, although not as dramaticaly as in the full model. However, unlike  the full model, the collinear phase is stable, although suppressed in the extended Heisenberg model. Furthermore, like the case for the full model, we found that the extended Heisenberg model also sustains a sizeable spin-liquid phase.  This is consistent with our recent Schwinger boson mean-field theory study of the effect of the third-nearest neighbour exchange terms on the magnetic properties of the anisotropic triangular lattice.\cite{Merino14}  

For the plaquette model in Fig. \ref{fig:phaseringquantum}(c), we find that the N\'{e}el phase is stable for all $J'/J \le 0.5$ up to $K/J \approx 0.1$, but further increasing ring exchange drives a quantum disordered phase for $J'/J \sim 0.5$. Additionally there is a spin liquid phase for all $K/J \ge 0$ for $J'/J > 3.75$. For $0.5 < J'/J < 3.75$ spiral order is robust to ring exchange, as predicted classically. 

Understanding the competing orders in both the extended Heisenberg and plaquette models is helpful in understanding the competing orders in the full model. The behaviour of the N\'{e}el phase in the full model (see Fig. \ref{fig:phaseringquantum}(a)) is as follows. For moderate values of $K/J$, the two-spin renormalization of the Heisenberg terms by ring exchange help stabilize N\'{e}el order while for large values of $K/J$ the four-spin terms from ring exchange render the N\'{e}el phase unstable since such terms  drive the development of a spin liquid phase. The change in order of the N\'{e}el-spin liquid phase transition is also consistent with this picture, as the phase transition remains second order in the extended Heisenberg model of all $K$, whereas this phase transition is always first order in the plaquette model. Similarly, the development of a large spin-liquid phase for large $K/J$ in the full model can be understood from the two-spin ring contributions wanting to drive a collinear ground state while the four-spin terms favor a spiral ground state.   

\begin{figure*}
\begin{center}
     \includegraphics[width=0.70\columnwidth, trim = 0mm 0mm 0mm 0mm, clip]{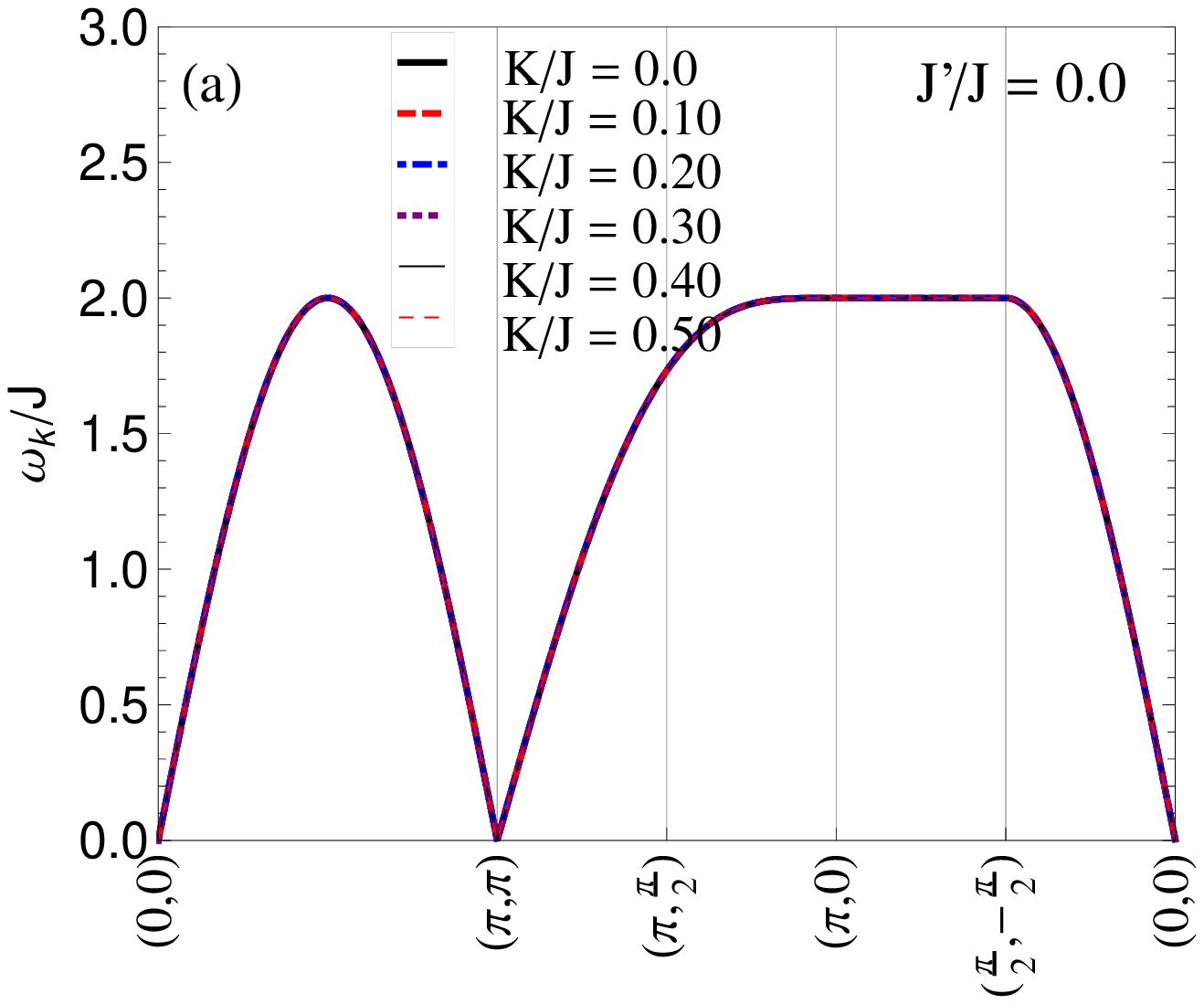}
	\includegraphics[width=0.70\columnwidth, trim = 0mm 0mm 0mm 0mm, clip]{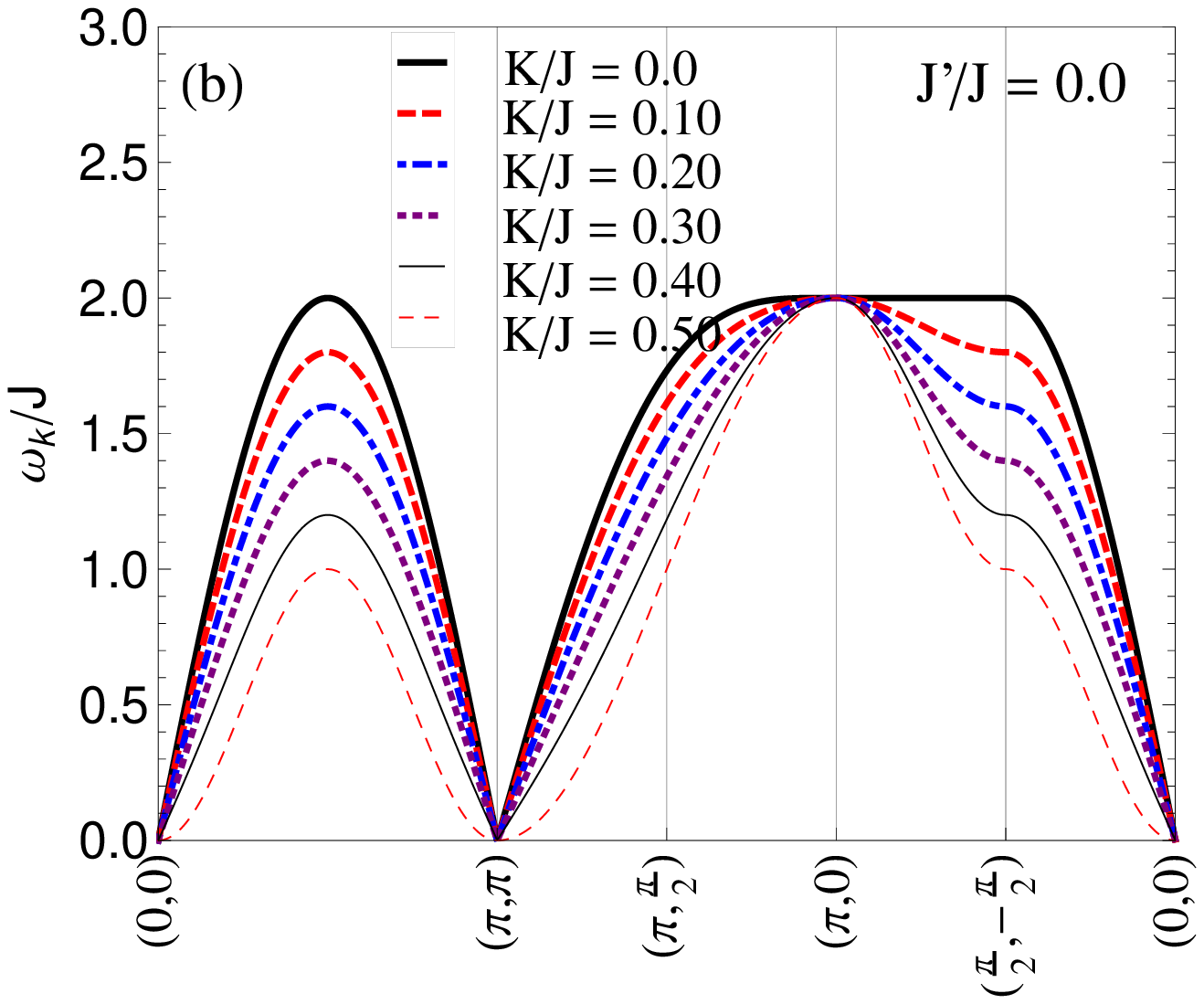}
   \end{center}
   \caption{(Color online) Comparison of spin-excitation spectrum on the square lattice obtained using (a) the full model $\hat{H} = \hat{H}_{2} + \hat{H}_{2}^{*} + \hat{H}_{4}$ and (b) the plaquette model $\hat{H} = \hat{H}_{2} + \hat{H}_{4}$.}
    \label{fig:squaredispersion}
\end{figure*}  

\subsection{Staggered Magnetization}

In Fig. \ref{fig:mstagmodels}(a) we present the staggered magnetization for the full model for the N\'{e}el and spiral phases as a function of $J'/J$ for several values of $K/J$. The strong dip in the magnetization in the region $0.5-0.8$ for the values of $K/J$ considered, indicates  a disordered intermediate phase. The nature of the ground-state cannot be determined from linear spin-wave theory and will have to be determined by more sophisticated techniques.

Examination of the staggered magnetization in this parameter region indicates that both the N\'{e}el-spin liquid and spiral-spin liquid phase boundaries are lines of second order phase transitions vanishing at a quantum critical point at $K=K'=0$, $J'/J=0.5$. This is consistent with what has previously been found in the $K=K'=0$ case \cite{Merino99,Trumper99}.   

In Fig. \ref{fig:mstagmodels}(b) and (c) we present the staggered magnetization for the extended Heisenberg and plaquette models respectively for the N\'{e}el and spiral phases as a function of $J'/J$ for several values of $K/J$. We observe that the two and four spin terms compete with each other with the former decreasing $m_{st}$ while the latter increasing $m_{st}$ with incrasing $K/J$. This results in the negligible change in $m_{st}$ in the N\'{e}el phase of the full model in Fig. \ref{fig:mstagmodels}(a). Furthermore, we observe the suppression of the spiral phase with increasing $K/J$ is driven purely by the two-spin terms.   

In Fig. \ref{fig:vanishmstagmodels} we present the staggered magnetization in the N\'{e}el phase for each of the three models. 
For the full and plaquette models we observe that for small values of $K/J$ the N\'{e}el order undegoes a second order phase transition to the spiral phase (Fig. \ref{fig:vanishmstagmodels}(a) and (c)), while for moderate $K/J$ the transition from N\'{e}el order to the spin-liquid phase is clearly first order. Therefore, the transition from enhancement to suppression of N\'{e}el order with increasing $K/J$ can be understood in terms of a quantum critical endpoint being reached where the quantum phase transition changes from second-order to first-order (marked by a red X in Fig. \ref{fig:phaseringquantum}(a)). For the extended Heisenberg model we observe that for large $K/J$ the transition from N\'{e}el to collinear order is first order.

\section{Spin-Excitation Spectra}

\label{sec:Excitations}

We begin our discussion of the excitation spectra by considering the N\'{e}el phase. In the N\'{e}el phase of the full model the antiferromagnetic exchange constants  (Eq. \eqref{eq:dressedexchange}) reduce to 
\begin{eqnarray}
\label{eq:Neelexchange}
J_{\hat{\bf x}} = J_{\hat{\bf y}} &=& J + 2K'\nonumber\\
J_{\hat{\bf x} + \hat{\bf y}} &=& J' + 8K'\nonumber\\
J_{\hat{\bf x} - \hat{\bf y}} &=& 0\nonumber\\
J_{2\hat{\bf x} + \hat{\bf y}} = J_{\hat{\bf x} + 2\hat{\bf y}} &=& 2K'
\end{eqnarray}
\noindent For the square lattice $K'=J'=0$, since $K' = J'K$, the spin-wave dispersion is independent of the ring exchange coupling $K$ in the full model,  Eq. \eqref{eq:bigHamiltonian}, 
\begin{eqnarray}
\label{eq:SquareDispersion}
\omega_{\bf k}^2 = \bigg (2J - J' + J(\cos(k_{x}) + \cos(k_{y})) + J'\cos(k_{x} + k_{y}) \bigg )\nonumber\\
\times\bigg (2J - J' - J(\cos(k_{x}) + \cos(k_{y})) + J'\cos(k_{x} + k_{y}) \bigg )\nonumber\\
\end{eqnarray}
\noindent We plot this in Fig. \ref{fig:squaredispersion}. 

In the plaquette model the antiferromagnetic exchange constants  (Eq. \eqref{eq:dressedexchange}) in the N\'{e}el phase are 
\begin{eqnarray}
\label{eq:NeelexchangeMajumdar}
J_{\hat{\bf x}} = J_{\hat{\bf y}} &=& J - 2K - K'\nonumber\\
J_{\hat{\bf x} + \hat{\bf y}} &=& J' + 4K' - K\nonumber\\
J_{\hat{\bf x} - \hat{\bf y}} &=& 0\nonumber\\
J_{2\hat{\bf x} + \hat{\bf y}} = J_{\hat{\bf x} + 2\hat{\bf y}} &=& 2K'
\end{eqnarray}
\noindent %This is a direct consequence of the neglect of the renormalization of the two-spin terms by ring exchange in the plaquette model. We note that the extended Heisenberg model is also dependent on ring exchange. 
However, the effect of ring exchange is opposite to that of the plaquette model, this results in the the near independence of the spin-excitation spectrum observed of the full model to ring-exchange observed in Fig. \ref{fig:squaredispersion}(a).

\begin{figure*}
\begin{center}
 \includegraphics[width=0.70\columnwidth, clip]{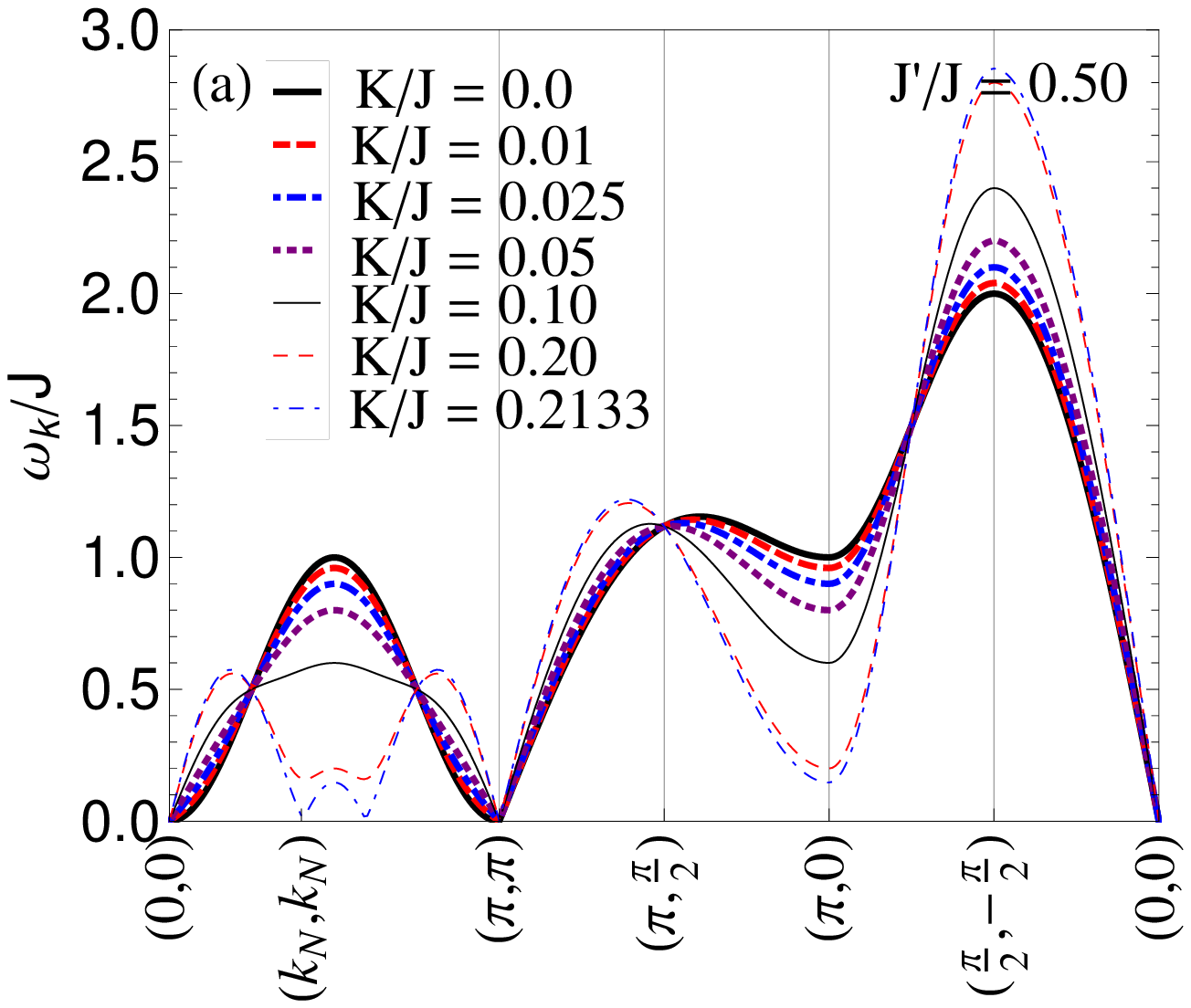}
 \includegraphics[width=0.70\columnwidth, clip]{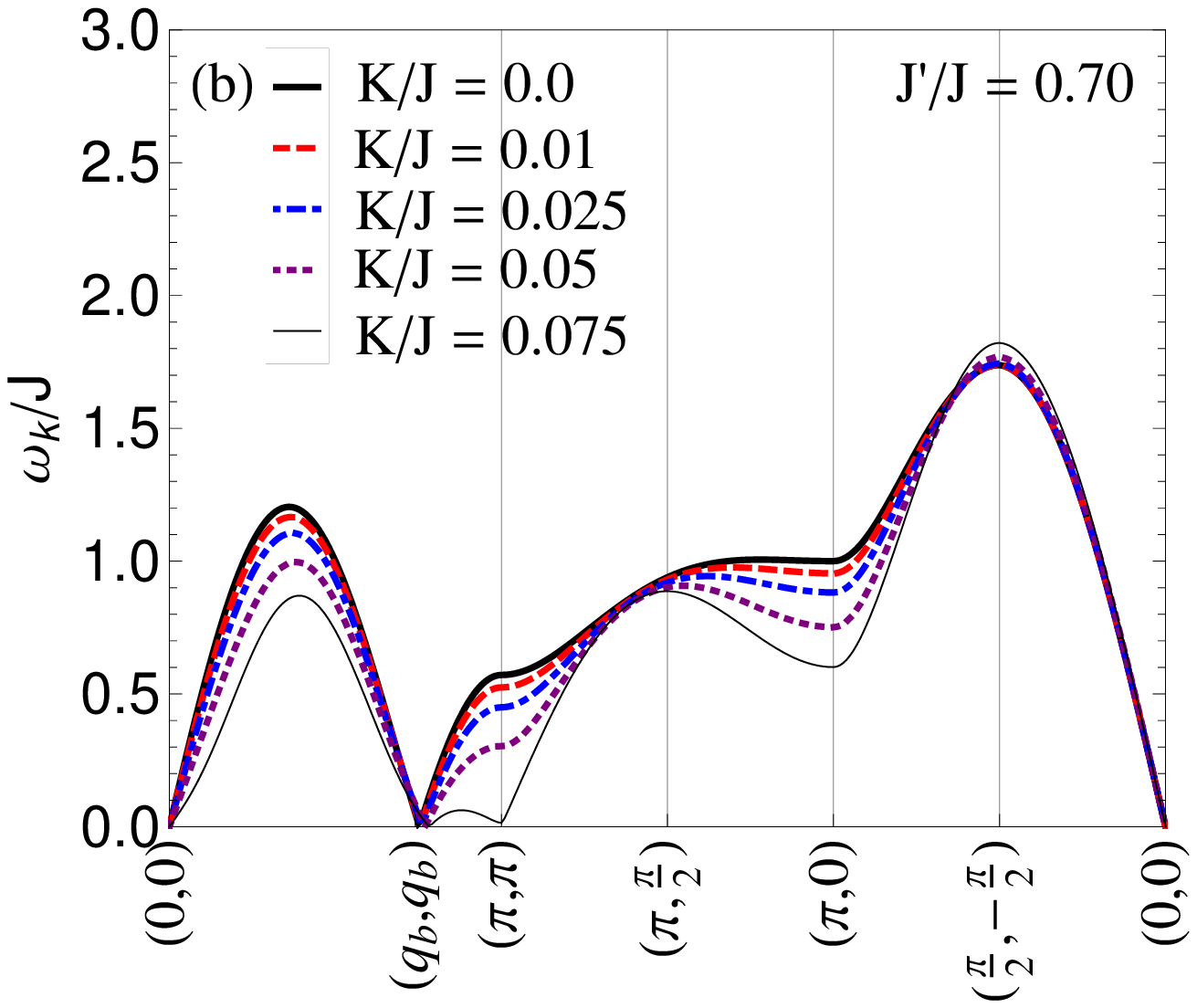}
\includegraphics[width=0.70\columnwidth, clip]{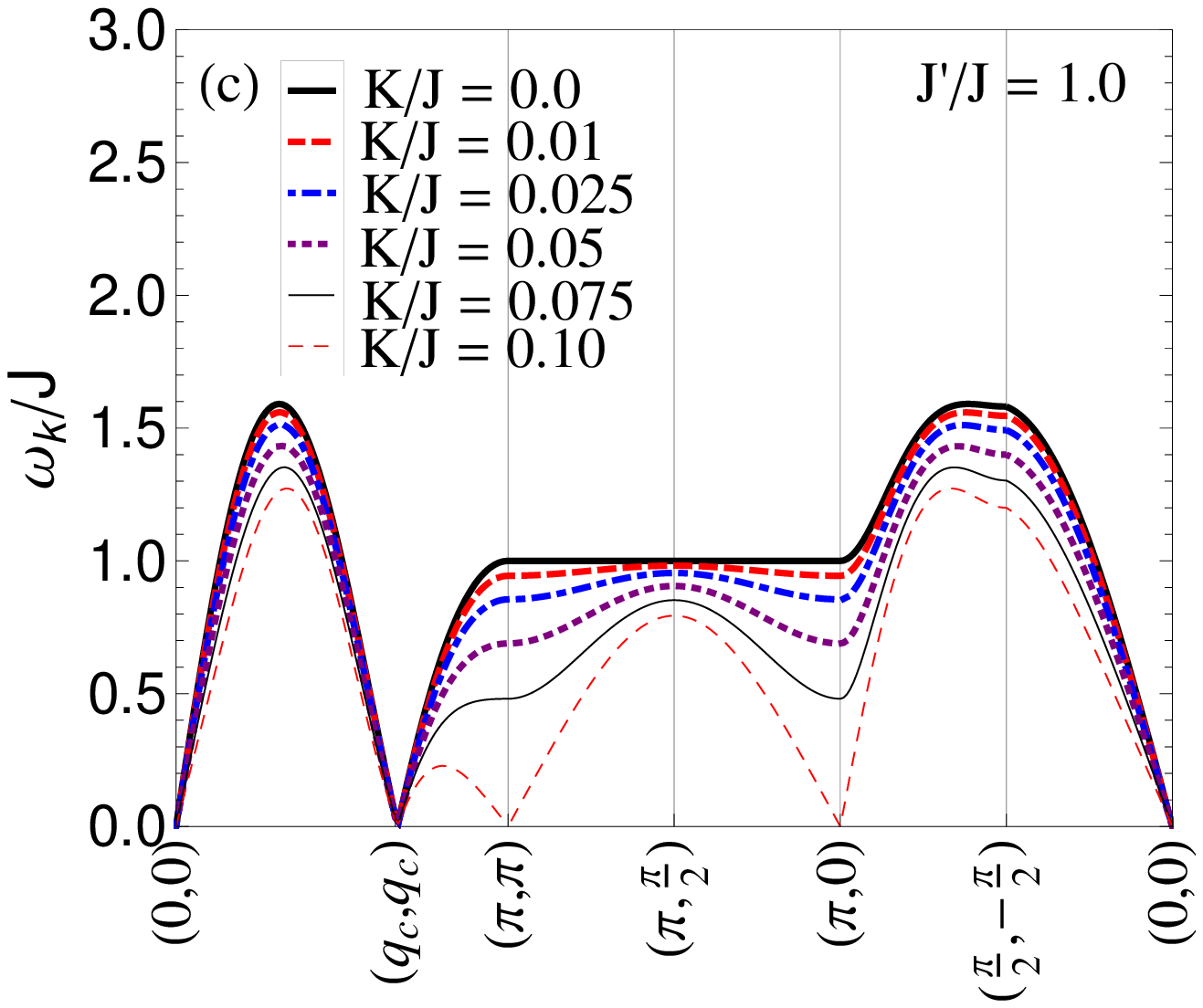}
\includegraphics[width=0.70\columnwidth, clip]{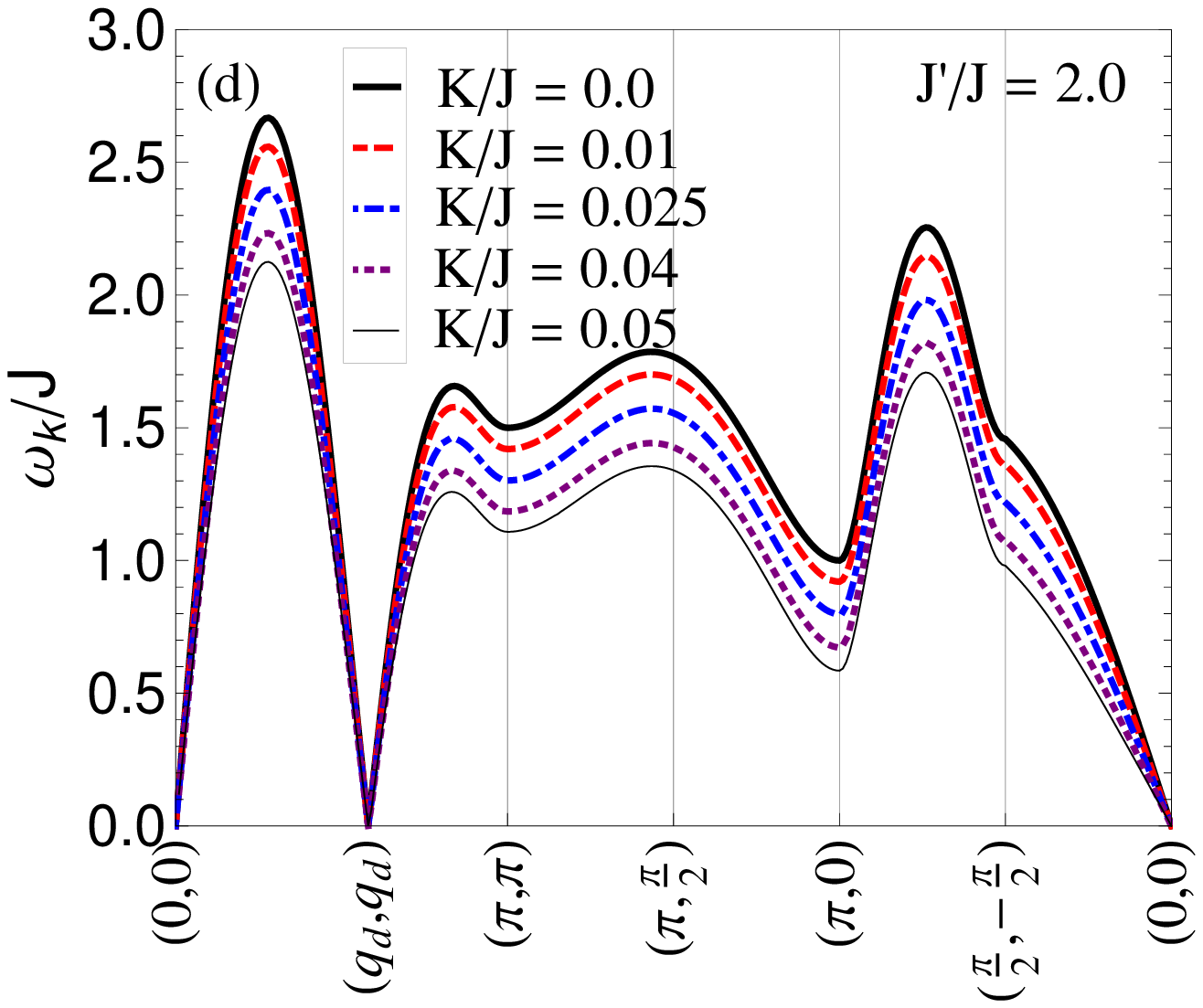}
    \end{center}
   \caption{(Color online) Spin-excitation spectra calculated for the full model for (a) the N\'{e}el phase with $J'/J = 0.5$ as well as for the commensurate spiral phase $J'/J = 0.7$ (b), $J'/J = 1.0$ (c) and $J'/J = 2.0$ (d). In (a)-(d) we mark spiral ordering vectors $k_{N}$, $q_{b}$, $q_{c}$ and $q_{d}$ with $k_{N} \approx 0.40\pi$, $q_{b} \approx 0.76\pi$, $q_{c} =\frac{2}{3}\pi$ and $q_{d} \approx 0.58\pi$.  In all cases ring exchange increases the competition between the different classical phases, which causes a dramatical softening of the dispersion at the competing ordering wavevectors in (a)-(c). Above the critical value of the ring exchange the dispersion becomes imaginary at these wavevectors - thus the competition between the different ordered phases is seen to be directly responsible for the quantum disordered phases.}
    \label{fig:dispersionring}
\end{figure*}
\begin{figure*}
 \begin{center}
      \includegraphics[width=0.7\columnwidth, trim = 0mm 0mm 0mm 0mm, clip]{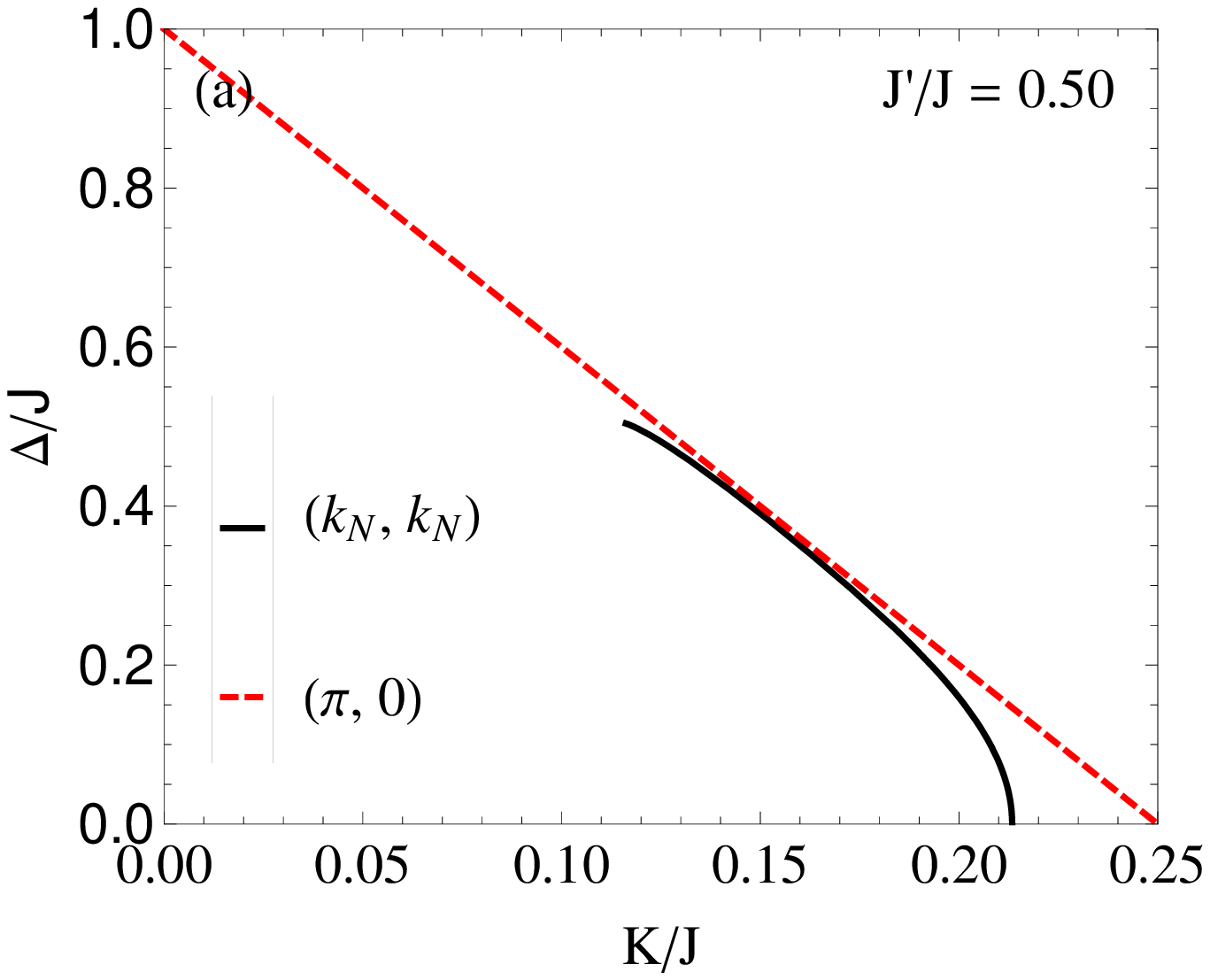}
	\includegraphics[width=0.7\columnwidth, trim = 0mm 0mm 0mm 0mm, clip]{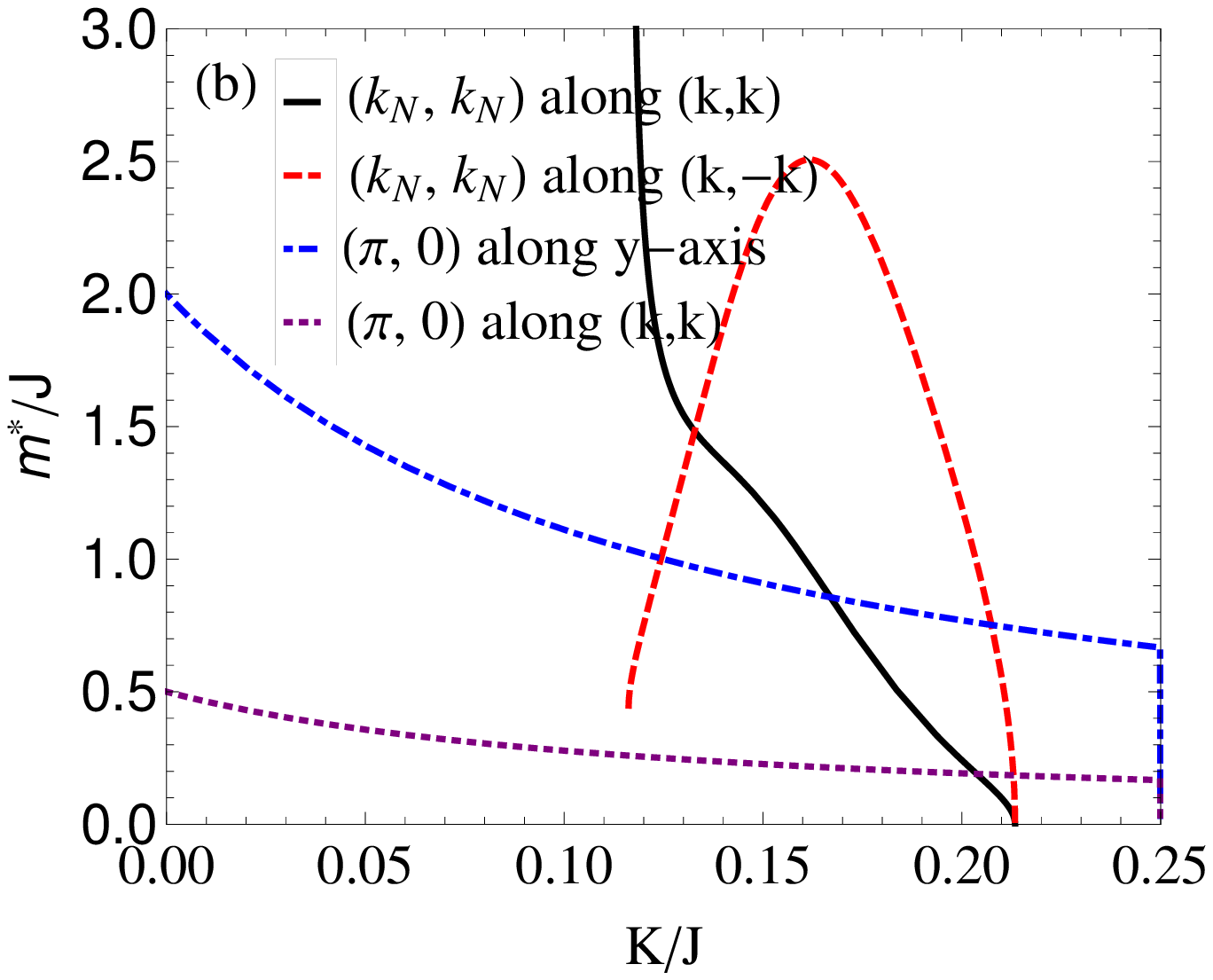}
    \end{center}
    \caption{(Color online) (a) spin-wave gap and (b) effective masses for the N\'{e}el phase of the full model as functions of $K/J$ for $J'/J = 0.5$. We consider both the ${\bf k} = {\bf k_{N}} = (k_{N}, k_{N})$ and ${\bf k} = (\pi, 0)$ modes as the compete with each other. In calculating the effective masses we calculate the derivative along the diagonal ${\bf k} = (k, k)$ and perpendicular ${\bf k} = (k_{N}/\sqrt{2} + k, k_{N}/\sqrt{2} - k)$ directions for the ${\bf k} = (k_{N}, k_{N})$ mode. These are shown by the black solid and red dashed curves respectively. Similarly for the ${\bf k} = (\pi, 0)$ mode we calculate the derivative along the boundary $(\pi, k)$ and $(k, \pi)$ directions as well as along the diagonal $(\pi/2 + k, \pi/2 - k)$ and $(\pi/2+k, -\pi/2 + k)$ directions. These are shown by the blue dot-dashed and purple dotted curves respectively.}
    \label{fig:effectivemass}
\end{figure*}
\begin{figure}
 \begin{center}
      \includegraphics[width=0.66\columnwidth, trim = 0mm 0mm 0mm 0mm, clip]{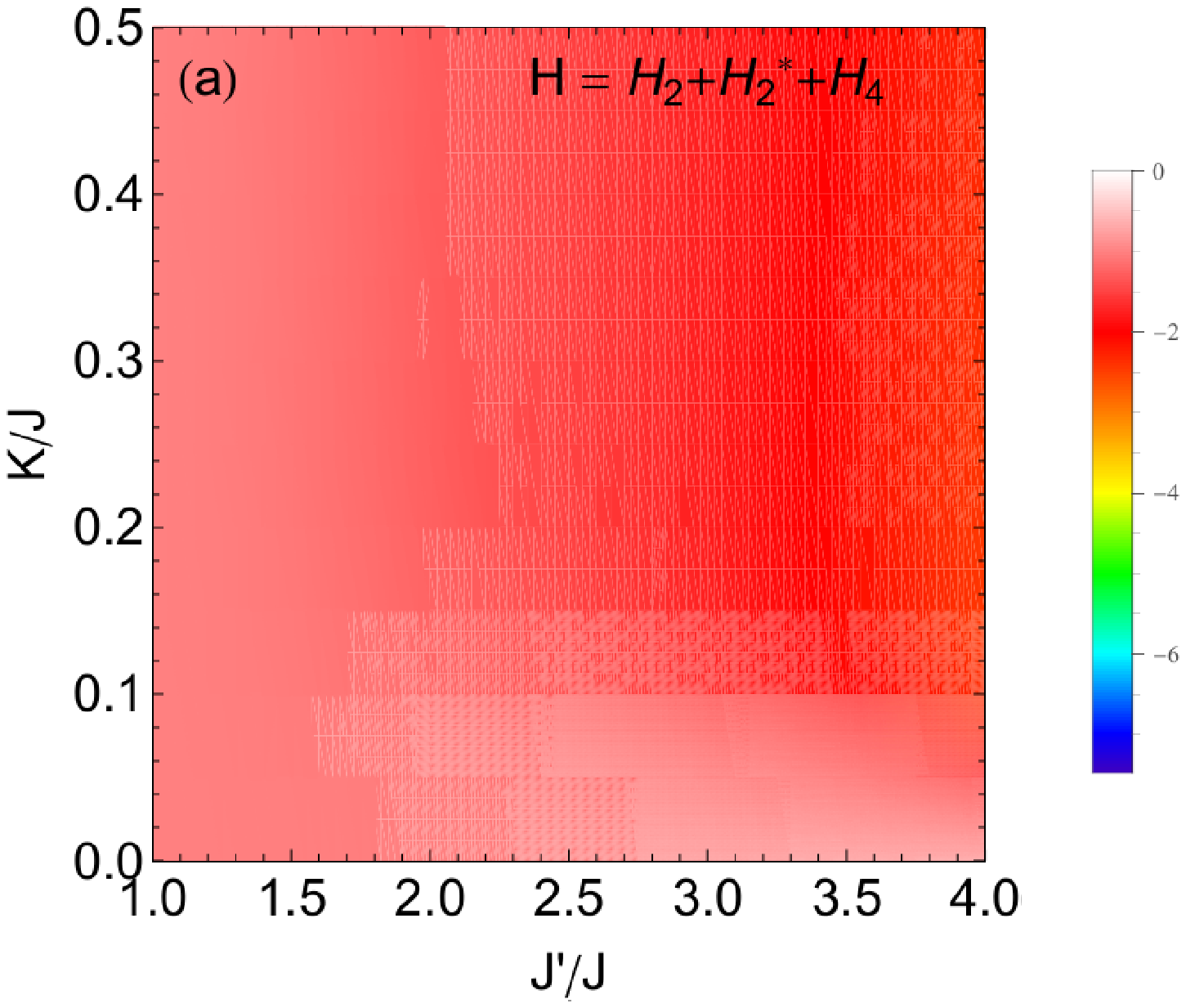}
      \includegraphics[width=0.66\columnwidth, trim = 0mm 0mm 0mm 0mm, clip]{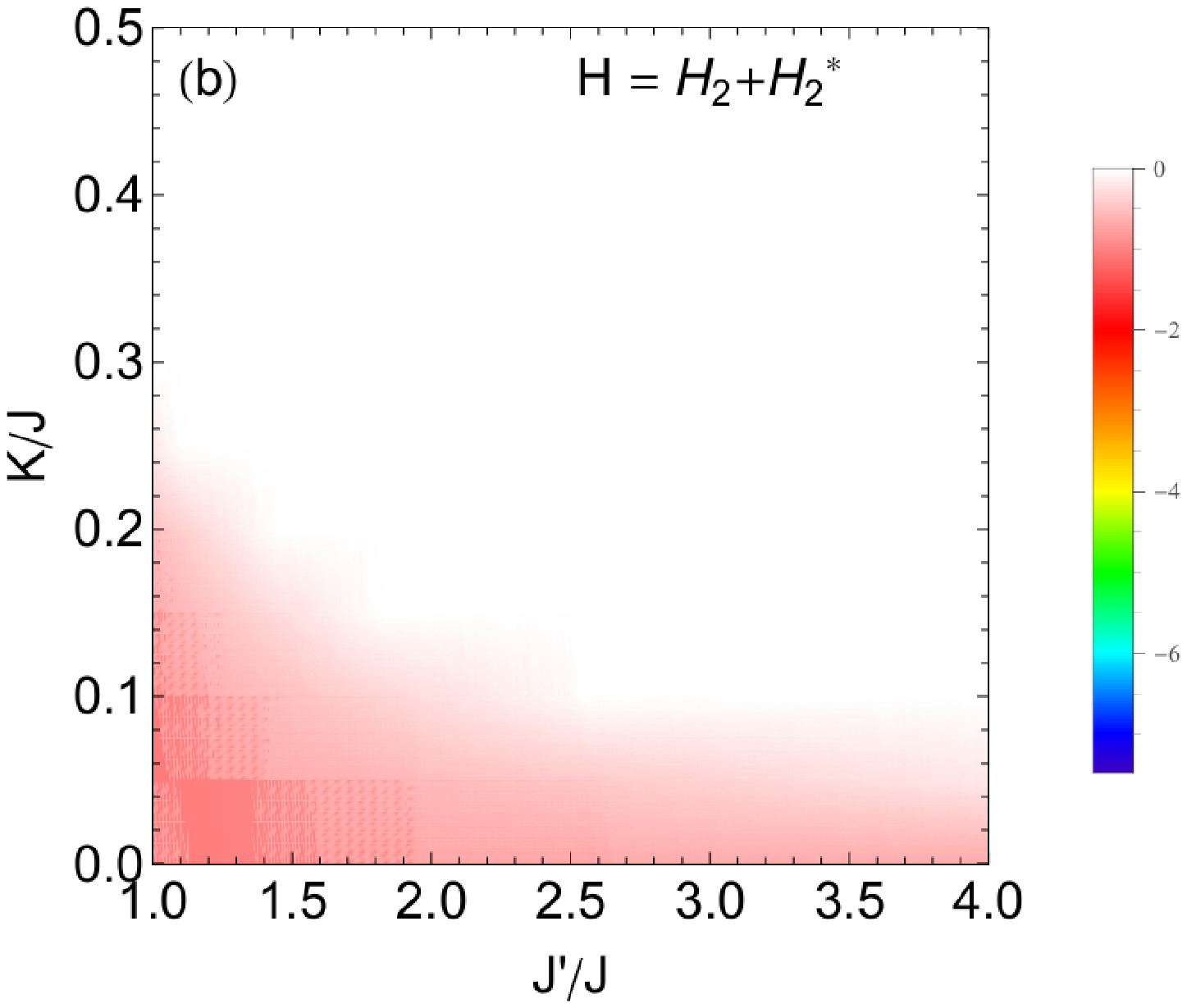}
	\includegraphics[width=0.66\columnwidth, trim = 0mm 0mm 0mm 0mm, clip]{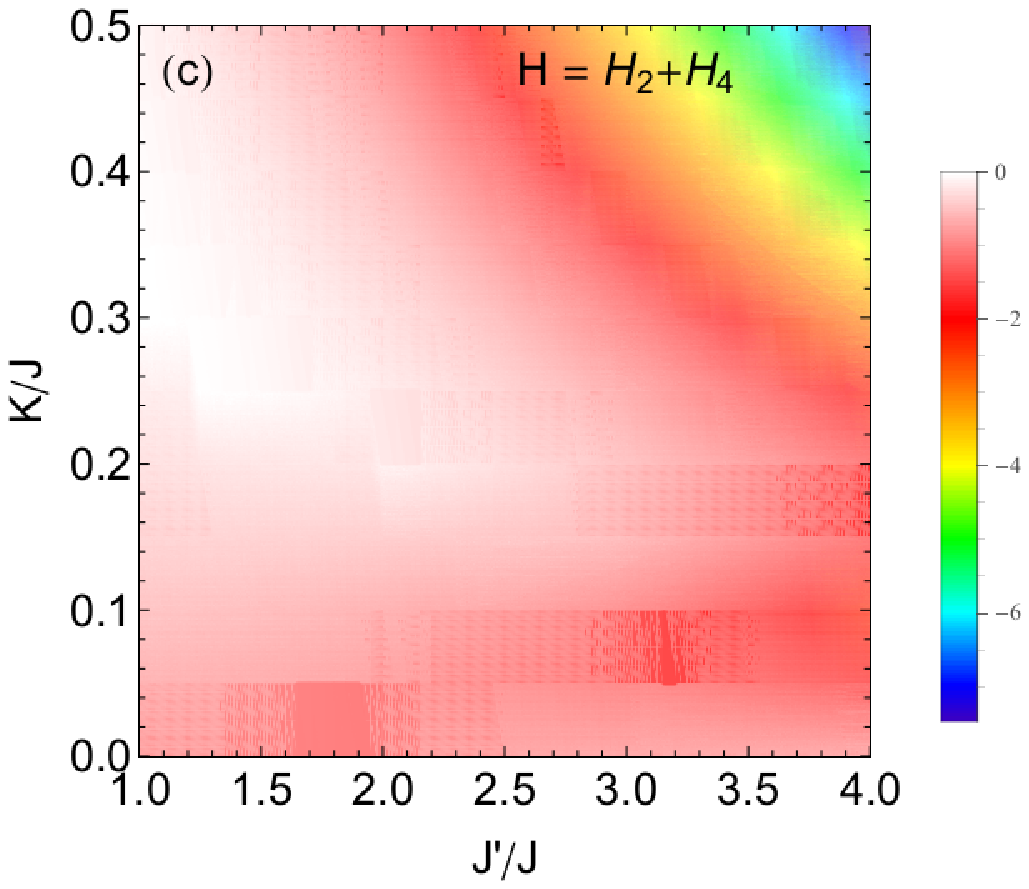}
\end{center}
    \caption{(Color online) Minimum value of $\omega_{\bf k}^2$ assuming ${\bf Q} = (\pi, 0)$ ordering. In (a) we use $\omega_{\bf k}^2$ obtained using the full model  (Eq. \eqref{eq:collinearspectra}); in (b) we use the spectrum calculated using the extended Heisenberg model  (Eq. \eqref{eq:collinearspectra2spin}); while in (c) we use the spectrum calculated using the plaquette model  (Eq. \eqref{eq:collinearspectra4spin}). Calculations were performed using LSWT.}
    \label{fig:collinearunstable}
\end{figure}

\subsection{Excitation Spectra For The Full Model}

In Fig. \ref{fig:dispersionring}(a) we plot the calculated spectra in the N\'{e}el phase for $J' = 0.5$. We find that increasing $K/J$ leads to the softening of the mode at ${\bf k}=(\pi,0)$ and another along the diagonal $\bf 0$-$\bm\pi$ direction, where ${\bf 0} = (0,0)$ and ${\bm \pi} = (\pi, \pi)$. The later is more physically significant as it drives the N\'{e}el-spin liquid transition. For sufficiently large $K/J$ local minima in $\omega_{\bf k}$ emerge at ${\bf k} = {\bf k_N} =(k_N,k_N)$ and ${\bf k} = {\bm\pi} - {\bf k_N}$ with
\begin{eqnarray}
\label{eq:kN}
k_{N} = \arccos \bigg (\frac{\sqrt{[\kappa_{1} - \sqrt{\kappa_{2}}]/K'^2}}{8\sqrt{3}} \bigg )
\end{eqnarray}
\noindent where 
\begin{eqnarray}
\label{eq:kNkappa12}
\kappa_{1} &=& J'^2 + 16K'(J' - J + 8K')\nonumber\\
\kappa_{2} &=& J'^4 + 32K'(J'^3 - J'^2(J-4K') - 4J'K'(J + 4K')\nonumber\\ 
&& + 2K'(J + 4K')^2)
\end{eqnarray}
\noindent As $K/J$ is further increased these minima deepen and eventually $\omega_{{\bf k_N}}=\omega_{{\bm\pi} - {\bf k_N}}$ becomes imaginary ($\omega_{\bf k_N}^2<0$). The softening of these modes can be understood by considering them to be massive modes with an ``effective mass"
\begin{eqnarray}
\label{eq:effectivemass}
\frac{1}{m^*} = \frac{\partial^2 \omega_{\bf k}}{\partial {\bf k}^2} \bigg |_{{\bf k}_{0}}
\end{eqnarray}

\noindent where ${\bf k}_{0}$ is the momentum of the mode of interest, i.e., the momentum where the local minimum occurs. In Fig \ref{fig:effectivemass} we plot the (a) gap and (b) effective mass of the ${\bf k} = (k_{N}, k_{N})$ and ${\bf k} = (\pi, 0)$ modes in the N\'{e}el phase for $J'/J = 0.5$ as functions of $K/J$. In calculating the effective masses we calculate the derivative along the ${\bf k} = (k, k)$ and ${\bf k} = (k_{N}/\sqrt{2} + k, k_{N}/\sqrt{2} - k)$ directions for the ${\bf k} = (k_{N}, k_{N})$ mode. Similarly for the ${\bf k} = (\pi, 0)$ mode we calculate the derivative along the $(\pi, k)$ and $(k, \pi)$ directions as well as along $(\pi/2 + k, \pi/2 - k)$ and $(\pi/2+k, -\pi/2 + k)$ directions.    

With increasing ring exchange we observe that the effective masses of both modes decrease. Eventually the ${\bf k} = (k_{N}, k_{N})$ mode becomes massless and Goldstone's theorem implies that the long-range order commensurate with that mode competes with the N\'{e}el phase. This leads to an instability of the N\'{e}el 	phase. This is a clear indication that N\'eel order has become unstable due to competition with the spiral phase. Explicit calculation shows that long range spiral order with ${\bf Q}={\bf k_N}$ or $\bm\pi-\bf k_N$ is also unstable in this parameter regime. Generally, the instability of ordered phases for $K/J > 0$ results from the competition between two (or more) classical orders. It is the competition between N\'eel $({\bf Q}=\bm\pi)$ and spiral $({\bf Q}={\bf k_N})$ order destroys the N\'{e}el long range order in the full model. This is very different from the mechanism for the vanishing of long range order at the quantum critical point in the $K=0$ limit which has been shown \cite{Merino99,Trumper99} to be due to the vanishing of the spin-wave velocity along $\bf 0$-$\bm \pi$ and can be observed in Fig. \ref{fig:dispersionring}(a).

In Fig. \ref{fig:dispersionring}(b)-(d) we plot the spectra in the spiral phase for $J' = 0.7$, $J'=J$ and $J'=2J$. In all three cases one can clearly observe the expected Goldstone modes at ${\bf k} = {\bf 0}$ and ${\bf k} = {\bf Q}=(q, q)$. In the spiral phase increasing $K/J$ induces softenings at ${\bf k} = {\bm \pi}$ and ${\bf k} = (\pi, 0)$, i.e., at the momenta of the Goldstone modes of the N\'{e}el and collinear phases respectively. For $J'/J = K'/K <1$ the mode softens most rapidly at ${\bf k}= \bm \pi$. For sufficiently large $K/J$ we find that $\omega_{\bm \pi}$ becomes imaginary (as $\omega_{\bm \pi}^2$ becomes negative) indicating that the competition with the N\'eel phase has destroyed the long range spiral order. For $J'/J = K'/K>1$ the mode softens most rapidly at ${\bf k}=(\pi, 0)$. For sufficiently large $K/J$ we find that $\omega_{(\pi, 0)}$ becomes imaginary indicating that the competition with the collinear phase has destroyed the long range spiral order. At $J'/J = K'/K = 1$ (Fig. \ref{fig:dispersionring}(c)) both the N\'eel and collinear phases compete with the spiral phases (as one would suspect from the classical phase diagram, Fig. \ref{fig:phasering}(a)) and the dispersion becomes imaginary at ${\bf k} = \bm \pi$ and ${\bf k} = (\pi,0)$ simultaneously and likewise their effective masses also vanish simultaneously.  A similar minimum at $(\pi,0)$ has been found from series expansions for the Heisenberg model on an anisotropic triangular lattice with no ring exchange due to recombination of particle-hole spinon pairs of momenta: $(\pi/2,\pi/2), (-\pi/2,-\pi/2)$ into magnons \cite{Zheng99, Fjaerestad07} in that case. 

It is also interesting to note that for $J'=J$, the spiral order is more robust to the disordering effects induced by the ring exchange than for any other value of $J'/J$ even classically. The effective two spin exchange couplings of Hamiltonian (\ref{eq:bigHamiltonian}) are  $\tilde{J'}=J'+K+4K'$ and $\tilde{J}=J+2K+3K'$ implying that $\tilde{J'}=\tilde{J}$  for all $K/J$ when $J'/J=K'/K=1$.  Furthermore, the strong geometrical frustration suppresses N\'{e}el and collinear phases thereby decreasing their ability to compete with spiral phase and drive an instability to the quantum spin liquid.

\subsection{Absence of Collinear Phase in Quantum Calculations}

We found that, in the parameter range covered by Fig. \ref{fig:phasering}, the collinear phase is classically stable (see Fig. \ref{fig:phasering}(a)). However, in the quantum phase diagram (see  Fig. \ref{fig:phaseringquantum}(a)) the collinear phase is always unstable as there is always some point (or, typically, area) of the Brillouin zone for which $\omega_{\bf k}^2<0$. To demonstrate that the spectrum is unstable in the full model we write down explicitly the expression for $\omega_{\bf k}^2$ assuming ${\bf Q} = (\pi, 0)$ ordering:
\begin{widetext}
\begin{eqnarray}
\label{eq:collinearspectra}
\omega_{\bf k}^2 = \bigg (J' + \bigg [J + 4(K + K') \bigg ]\cos(k_{y}) \bigg )^2 - \bigg (\cos(k_{x})\bigg [J + 2K' + 4K\cos(k_{y}) \bigg ] + J'\cos(k_{x}+k_{y}) + 2K'\cos(k_{x}+2k_{y}) \bigg )^2
\end{eqnarray}
\end{widetext}
\noindent In Fig. \ref{fig:collinearunstable}(a) we plot the minimum values (with respect to $\bf k$) of  Eq. \eqref{eq:collinearspectra} and notice that $\omega_{\bf k}^2 \leq -1$ for the entire parameter space considered. Therefore we conclude that competition with other classical phases means that the collinear phase is not stable in our calculations.

An important question to answer is why is the collinear phase so fragile in the quantum calculations for the full model. To address this we consider explicit expressions for the spin-excitation spectrum in the collinear phase for the extended Heisenberg
\begin{widetext}
\begin{eqnarray}
\label{eq:collinearspectra2spin}
\omega_{\bf k, eH}^2 &=& \bigg (J' + 2K + 4K' + (J + 2K + 3K')\cos(k_{y}) + K'\cos(2k_{x} + k_{y}) \bigg )^2 - \bigg ((J + 2K + 3K')\cos(k_{x}) + K\cos(k_{x}-k_{y})\nonumber\\ 
&& + (J' + K + 4K')\cos(k_{x}+k_{y}) + K'\cos(k_{x} + 2k_{y}) \bigg )^2,
\end{eqnarray}
\noindent and  plaquette models
\begin{eqnarray}
\label{eq:collinearspectra4spin}
\omega_{\bf k, P}^2 &=& \bigg (J' - 2K - 4K' + (J + 2K + K')\cos(k_{y}) -  K'\cos(2k_{x} + k_{y}) \bigg )^2 - \bigg ((J - 2K - K')\cos(k_{x}) + K\cos(k_{x}-k_{y})\nonumber\\  
&& + (J' + K -4K')\cos(k_{x}+k_{y}) + K'\cos(k_{x} + 2k_{y}) \bigg )^2
\end{eqnarray}
\end{widetext}
\noindent We plot the values of the minima (with respect to $\bf k$) of  Eqs. \eqref{eq:collinearspectra2spin} and   Eq. \eqref{eq:collinearspectra4spin} in Figs. \ref{fig:collinearunstable}(b) and (c). There is a broad region of the phase diagram where the collinear phase is stable for the extended Heisenberg model, but is  unstable in the entire parameter space considered for the plaquette model. Therefore, we conclude that it is the four-site ring exchange terms, rather than the renormalization of the exchange couplings that destabilizes the collinear phase. 

\begin{figure}
 \begin{center}
      \includegraphics[width=0.8\columnwidth, trim = 0mm 0mm 0mm 0mm, clip]{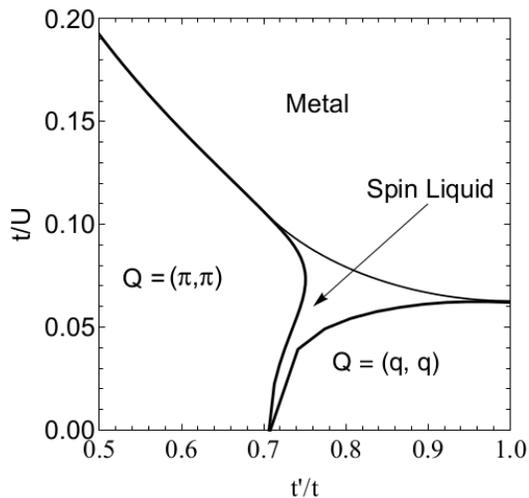}
\end{center}
    \caption{(Color online) Qualitative sketch of a proposed phase diagram for the Hubbard model on the anisotropic triangular lattice with ring exchange based on the LSWT calculations reported here and electronic structure calculations \cite{Scriven12}.
}
    \label{fig:proposed}
\end{figure}

One might wonder whether the instability of the collinear phase at harmonic order (leading order in $1/S$) in the spin wave calculations is due to the spin waves in this phase not being properly described by this level of treatment. Hartree-Fock corrections, coming from the spin-wave interaction terms could stabilize the collinear phase. We have performed such a calculation but we find that the collinear phase is also unstable for the full model with points (or regions) of the Brillouin zone where the renormalized spin excitation spectrum is imaginary. We will discuss this calculation in detail in a forthcoming manuscript, where we will calculate the full quantum phase diagram using Hartree-Fock mean-field theory. %We have also recently found that the collinear phase is stable in a related Heisenberg model with couplings to second and third nearest neighbours on the triangular lattice using Schwinger boson mean-field theory \cite{Merino14}. 

\section{Comparison with Organics}

\label{sec:Organics}

So far we have limited the discussion to the spin degrees of freedom only. However, in the materials of interest, particularly the organic charge transfer salts, the charge degrees of freedom eventually become important and a Mott transition occurs under pressure.  For $J'=J$ (120$^\circ$) spiral order is found for $K/J\lesssim0.1$. To lowest order $U/t=\sqrt{20J/K}$, which would suggest that the spiral-spin liquid transition occurs at $U/t\simeq14$ which is in good agreement with the previous calculations of Motrunich \cite{Motrunich05} and Yang {\it et al}. \cite{Yang10} for the isotropic triangular lattice model. This is also close to the estimated value of the critical ratio of $U/t$ for the Mott transition on the triangular lattice \cite{Merino06}. This suggests that for $J'\sim J$ there is a direct transition from a spiral ordered Mott insulator to a metal as pressure is increased, which is believed to decrease $U/t$. \cite{Scriven09a, Scriven09b, Powell11, Kanoda10} This is consistent with the observation\cite{Scriven12} that organic charge transfer salts with $t'\simeq t$ undergo a Mott transition directly from a magnetically ordered phase to a superconducting/metallic phase. Whereas salts with $J'/J\sim0.8$ display spin liquid (or other exotic quantum) phases.

In Fig. \ref{fig:proposed} we present a qualitative sketch of a proposed phase diagram for the Hubbard model on the anisotropic triangular lattice with ring exchange. The boundary lines for the N\'{e}el and spiral ordered phases are based on the linear spin-wave theory for the full model, reported above. It is well known that the perfect nesting of the square lattice means that is insulating for arbitrarily small $U/t$. For $t'=t$ it is found, numerically, that the Mott transition occurs at around $U/t=10-15$, depending on the method used.\cite{Powell11,Merino06,RVB3, Watanabe06, Sahebsara06, Kyung06, RVB, Chen13, Tocchio13} The metal-insulator transition line in Fig. \ref{fig:proposed} is simply a smooth curve joining these points.
Nevertheless this simple analysis suggests that there will be a region of the phase diagram in the full model where the magnetic orderings compete strongly enough to induce a stable spin-liquid region. Comparing the observed phase diagrams of the $\kappa$-(BEDT-TTF)${_2}X$ and $Y$[Pd(dmit)$_2$]$_2$ salts to this picture and taking into account the frustration ($J'/J$) estimated from first principles calculations \cite{Scriven12} one finds that this is consistent with what is observed experimentally.

\section{Conclusions}

\label{sec:Conclusions}

In this paper we have shown that the competition between different long range order states creates a quantum disordered phase in the anisotropic triangular lattice Heisenberg model with ring exchange even at the semiclassical level. Analysis of the spin wave spectra show that the spin liquid state is a consequence of competition between classical ordered states. %Such a competition between classical ordered states can be understood by considering effective masses of different modes in the Brillouin zone as well as the different ring exchange terms which contribute to the model. 
Thus we conclude that the interplay of ring exchange and geometrical frustration is responsible for the spin liquid state found. Our results are relevant to weak Mott insulators {\it i.e.} insulators lying close to the insulator-to-metal transition so that ring exchange is relevant.

A future challenge is understanding ring exchange effects on two-dimensional metals close to the Mott transition which may lead to exotic non-Fermi liquid $d$-wave \cite{Fisher12} phases.

\section{Acknowledgements}

This work was funded in part by the Australian Research Council under the Discovery (DP1093224), Future (FT130100161) and and QEII (DP0878523) schemes. J. M. acknowledges financial support from MINECO (MAT2012-37263-C02-01).


\begin{thebibliography}{99}

\expandafter\ifx\csname natexlab\endcsname\relax\def\natexlab#1{#1}\fi

\expandafter\ifx\csname bibnamefont\endcsname\relax

  \def\bibnamefont#1{#1}\fi

\expandafter\ifx\csname bibfnamefont\endcsname\relax

  \def\bibfnamefont#1{#1}\fi

\expandafter\ifx\csname citenamefont\endcsname\relax

  \def\citenamefont#1{#1}\fi

\expandafter\ifx\csname url\endcsname\relax

  \def\url#1{\texttt{#1}}\fi

\expandafter\ifx\csname urlprefix\endcsname\relax\def\urlprefix{URL }\fi

\providecommand{\bibinfo}[2]{#2}

\providecommand{\eprint}[2][]{\url{#2}}


\bibitem{Balents10}

L. Balents, Nature {\bf464}, 199 (2010).


\bibitem{Normand09}

B. Normand, Comtemp. Phys. {\bf50}, 533 (2009).


\bibitem{Powell11}

B. J. Powell and R. H. Mckenzie, Rep. Prog. Phys. {\bf 74}, 056501 (2011).


\bibitem{Lee08}

P. A. Lee, Science {\bf321}, 1306 (2008).


\bibitem{Shimizu03}

Y. Shimizu, K. Miyagawa, K. Kanoda, M. Maesato, and G. Saito, Phys. Rev. Lett. {\bf 91}, 107001 (2003).


\bibitem{Yamashita08}

S. Yamashita, Y. Nakazawa, M. Oguni, Y. Oshima, H. Nojiri, Y. Shimizu, K. Miygawa and K. Kanoda, Nature Phys. {\bf4}, 459 (2008).


\bibitem{Yamashita09} 

M. Yamashita, N. Nakata, Y. Kasahara, T. Sasaki, N. Yoneyama, N. Kobayashi, S. Fujimoto, T. Shibauchi and Y. Matsuda, Nature Phys. {\bf5}, 44 (2009).

\bibitem{Kanoda10}

K. Kanoda and R. Kato, Annu. Rev. Condens. Matter Phys. {\bf2}, 167 (2010).

\bibitem{Yamashita10} 

M. Yamashita, N. Nakata, Y. Senshu, M. Nagata, H. M. Yamamoto, R. Kato, T. Shibauchi and Y. Matsuda, Science {\bf328}, 1246 (2010).


\bibitem{Itou10}

T. Itou, A. Oyamada, S. Maegawa and R. Kato, Nature Phys. {\bf6}, 673 (2010).


\bibitem{Shimizu07} 

Y. Shimizu, H. Akimoto, H. Tsujii, A. Tajima, and R. Kato, J. Phys. Condens. Matter {\bf 19}, 145240 (2007).


\bibitem{Fjaerestad07}

J. O. Fj{\ae}restad, W. Zheng, R. R. P. Singh, R. H. McKenzie, and R. Coldea, Phys. Rev. B, {\bf75}, 174447 (2007).


\bibitem{Shirata12}

Y. Shirata, H. Tanaka, A. Matsuo, and K. Kindo, Phys. Rev. Lett. {\bf 108}, 057205 (2012). 


\bibitem{Zhou11}

H. D. Zhou, E. S. Choi, G. Li, L. Balicas, C. R. Wiebe, Y. Qiu, J. R. D. Copley, and J. S. Gardner, Phys. Rev. Lett, {\bf 106}, 147204 (2011).


\bibitem{RVB3}

J. Liu and J. Schmalian, and N. Trivedi, Phys. Rev. Lett  {\bf94}, 127003 (2005). 


\bibitem{Watanabe06}

T. Watanabe, H. Yokoyama, Y. Tanaka, and J. Inoue, J. Phys. Soc. Jpn. {\bf 75}, 074707 (2006).


\bibitem{Sahebsara06}

P. Sahebsara and D. Senechal, Phys. Rev. Lett. {\bf 97}, 257004 (2006).


\bibitem{Kyung06}

B. Kyung and A. M. S. Tremblay, Phys. Rev. Lett. {\bf 97}, 046402 (2006).


\bibitem{RVB}

B. J. Powell and R. H. McKenzie, Phys. Rev. Lett {\bf98} 027005 (2007).


\bibitem{Chen13}

K. S. Chen, Z. Y. Meng, U. Yu, S. Yang, M. Jarrell, and J. Moreno, Phys. Rev. B {\bf 88}, 041103(R) (2013).


\bibitem{Tocchio13}

L. F. Tocchio, H. Feldner, F. Becca, R. Valent\'{i}, and C. Gros, Phys. Rev. B {\bf 87}, 035143 (2013).


\bibitem{Scriven12}

E. P. Scriven and B. J. Powell, Phys. Rev . Lett. {\bf 109}, 097206 (2012).


\bibitem{Nakamura09}

K. Nakamura, Y. Yoshimoto, T. Kosugi, R. Arita and M. Imada, J. Phys. Soc. Jpn. {\bf78} 083710 (2009). 


\bibitem{Kandpal09}

H. C. Kandpal, I. Opahle, Y.-Z. Zhang, H. O. Jeschke, and R. Valenti, Phys. Rev. Lett. {\bf 103}, 067004 (2009).


\bibitem{Nakamura12}

K. Nakamura, Y. Yoshimoto and M. Imada, Phys. Rev. B {\bf86}, 205117 (2012).


\bibitem{Tsumuraya13}

T. Tsumuraya, H. Seo, M. Tsuchiizu, R. Kato, T. Miyazaki, J. Phys. Soc. Jpn., {\bf82}, 033709 (2013).


\bibitem{Susuki13}

T. Susuki, N. Kurita, T. Tanaka, H. Nojiri, A. Matsuo, K. Kindo, and H. Tanaka, Phys. Rev. Lett {\bf 110}, 267201 (2013).


\bibitem{Powell06}

B. J. Powell and R. H. Mckenzie, J. Phys.: Condens. Matter {\bf 18}, R827 (2006).


\bibitem{Yamaura04}

J. I. Yamaura, A. Nakao, and R. Kato, J. Phys. Soc. Jpn. {\bf 73}, 976 (2004).


\bibitem{MacDonald88}

A. H. MacDonald, S. M. Girvin, and D. Yoshioka, Phys. Rev. B {\bf 37}, 9753 (1988).


\bibitem{Balents03}

L. Balents and A. Paramekanti, Phys. Rev. B {\bf 67}, 134427 (2003).


\bibitem{Motrunich05}

O. I. Motrunich, Phys. Rev. B {\bf 72} 045105 (2005).


\bibitem{Yang10}

H. Y. Yang, A. M. Lauchli, F. Mila and K. P. Schmidt, Phys. Rev. Lett. {\bf 105}, 267204 (2010).


\bibitem{fn1}

Note in the convention followed in this paper $J$ is a factor of two larger than in Ref. \onlinecite{Motrunich05}. 




\bibitem{Roger83}

M. Roger, J. H. Hetherington, and J. M. Delrieu, Rev. Mod. Phys. {\bf 55}, 1 (1983).


\bibitem{Thouless65}

D. J. Thouless, Proc. Phys. Soc. {\bf 86}, 893 (1965).

\bibitem{Franco86}

H. Franco, R. E. Rapp, and H. Godfrin, Phys. Rev. Lett. {\bf 57}, 1161 (1986).


\bibitem{Godfrin88}

H. Godfrin, R. R. Ruel, and D. D. Osheroff, Phys. Rev. Lett. {\bf 60}, 305 (1988).


\bibitem{Misguich98}

G. Misguich, B. Bernu, C. Lhuillier, and C. Waldtmann, Phys. Rev. Lett {\bf 81}, 1098 (1998). 


\bibitem{Misguich99} 

G. Misguich, C. Lhuillier, B. Bernu, and C. Waldtmann, Phys. Rev. B {\bf 60}, 1064 (1999).


\bibitem{Roger98}

M. Roger, C. B\"aerle, Yu. M. Bunkov, A. S. Chen, and H. Godfrin,  Phys. Rev. Lett. {\bf 80}, 1308. (1998)


\bibitem{Kubo97}

K. Kubo and T. Momoi, Z. Phys. B {\bf 103}, 485 (1997).


\bibitem{Momoi97}

T. Momoi, K. Kubo and K. Niki, Phys. Rev. Lett. {\bf 79}, 2081 (1997).


\bibitem{Kubo98}

K. Kubo, H. Sakamoto, T. Momoi, and K. Niki, J. Low. Temp. Phys. {\bf 111}, 583 (1998).


\bibitem{Momoi99}

T. Momoi, H. Sakamoto, and K. Kubo, Phys. Rev. B {\bf 59}, 9491 (1999).


\bibitem{Kubo03}

K. Kubo and T. Momoi, Physica B {\bf 142}, 329 (2003).


\bibitem{Yasuda06}

C. Yasuda, D. Kinouchi, and K. Kubo, J. Phys. Soc. Jpn. {\bf 75}, 104705 (2006).


\bibitem{Sugai90}

S. Sugai, M. Sato, T. Kobayashi, J. Akimitsu, T. Ito, H. Takagi, S. Uchida, S. Hosoya, T. Kajitani, and T. Fukuda, Phys. Rev. B {\bf 42}, 1045 (1990).


\bibitem{Coldea01}

R. Coldea, S. M. Hayden, G. Aeppli, T. G. Perring, C. D. Frost, T. E. Mason, S. W. Cheong, and Z. Fisk, Phys. Rev. Lett. {\bf 86}, 5377 (2001).


\bibitem{Nunner02}

T. S. Nunner, P. Brune, T. Kopp, M. Windt, and Markus Gr\"{u}ninger, Phys. Rev. B {\bf 66}, 180404(R) 2002.


\bibitem{Anderson74}

P. W. Anderson, Mat. Res. Bull {\bf8} 153 (1973); .P. Fazekas and P. W. Anderson, Philos. Mag. {\bf30} 423 (1974).


\bibitem{Sindzingre94}

P. Sindzingre, P. Lecheminant  and C. Lhuillier  Phys. Rev. B {\bf50}, 3108 (1994).


\bibitem{Merino99}

J. Merino, R. H. McKenzie, J. B. Marston, and C. H. Chung, J. Phys. Condens. Matter {\bf 11}, 2965 (1999).


\bibitem{Trumper99}

A. E. Trumper, Phys. Rev. B {\bf60}, 2987 (1999).


\bibitem{Hauke13}

P. Hauke, Phys. Rev. B {\bf87}, 014415 (2013).


\bibitem{Zheng99}

Weihong Zheng , R. H. McKenzie, and R. R. P. Singh, Phys. Rev. B {\bf59}, 14367 (1999).


\bibitem{Bishop09}

R. F. Bishop, P. H. Y. Li, D. J. J. Farnell, and C. E. Campbell, Phys. Rev. B {\bf79}, 174405 (2009).


\bibitem{ChungJPCM01}

C. H. Chung, J. B. Marston, and R. H. McKenzie, J. Phys.: Cond. Matt. {\bf13}, 5159 (2001).


\bibitem{Monte}

S. Yunoki and S. Sorella, Phys. Rev. B {\bf74}, 014408 (2006); D. Heidarian, S. Sorella, and F. Becca, Phys. Rev. B {\bf80}, 012404 (2009).


\bibitem{RVB1}

B. J. Powell and R. H. McKenzie, Phys. Rev. Lett {\bf94}, 047004 (2005). 


\bibitem{RVB2}

J. Y. Gan, Y. Chen, Z. B. Su, and F. C. Zhang, Phys. Rev. Lett {\bf94}, 067005 (2005). 


\bibitem{RVB4}

Y. Hayashi and M. Ogata, J. Phys. Soc. Japan {\bf76}, 053705 (2007).


\bibitem{Reuther11}

J. Reuther and R. Thomale, Phys. Rev. B {\bf 83}, 024402 (2011).


\bibitem{Rau11}

J. G. Rau and H.-Y. Kee, Phys. Rev. Lett. {\bf106}, 056405 (2011).


\bibitem{Starykh07}

O. A. Starykh and L. Balents, Phys. Rev. Lett. {\bf98} 077205 (2007).


\bibitem{Weng06}

M. Q. Weng, D. N. Sheng, Z. Y. Weng, and R. J. Bursill, Phys. Rev. B {\bf74} 012407 (2006); A Weichselbaum and S. R. White Phys. Rev. B {\bf84}, 245130 (2011).


\bibitem{Sheng09}

 D. N. Sheng, O. I. Motrunich and M. P. A. Fisher, Phys. Rev. B {\bf79} 205112 (2009).
 

\bibitem{Block11}

M. S. Block, D. N. Sheng, O. I. Motrunich, and M. P. A. Fisher, Phys. Rev. Lett. {\bf106}, 157202 (2011).


\bibitem{Li10} 

H. Li. R. T. Clay and S. Muzumdar, J. Phys.: Condens. Matter {\bf22}, 272201 (2010).


\bibitem{Merino14}

J. Merino, M. Holt, B. J. Powell, arXiv:1402.3463


\bibitem{Majumdar12} 

K. Majumdar, D. Furton and G. S. Uhrig, Phys. Rev. B {\bf 85}, 144420 (2012).































\bibitem{Villain}

J. Villain, J. Physique, {\bf 38} 26 (1977); J. Villain, R. Bidaux, J. P. Carton, and R. Conte, J. Physique, {\bf 41}, 1263 (1980). 


\bibitem{Chandra90}

P. Chandra, P. Coleman and A. I. Larkin, Phys. Rev. Lett. {\bf 64}, 88 (1990); P. Chandra, P. Coleman and A. I. Larkin J. Phys. Condens. Matter {\bf 2}, 7933 (1990).


\bibitem{Chubukov92}

A. V. Chubukov and T. Jolicoeur, Phys. Rev. B {\bf 46}, 11137 (1992).











\bibitem{Miyake92}

S. Miyake, J. Phys. Soc. Jpn. {\bf 61}, 983 (1992).


\bibitem{Singh92}

R. R. P. Singh and D. A. Huse, Phys. Rev. Lett. {\bf 68}, 1766 (1992).


\bibitem{Sheng92}

Q. Sheng and C. L. Henley, J. Phys. Condens. Matt. {\bf 4}, 2937 (1992).


\bibitem{Hauke11}

P. Hauke, T. Roscilde, V. Murg, J. Cirac, and R. Schmied, New J. Phys. {\bf 13}, 075017 (2011).


\bibitem{Merino06} 

J. Merino, B. J. Powell and R. H. McKenzie, Phys. Rev. B {\bf 73}, 235107 (2006);
A. Liebsch, H. Ishida, and J. Merino, Phys. Rev. B {\bf 79}, 195108 (2009).


\bibitem{Scriven09a}

E. Scriven and B. J. Powell, J. Chem. Phys. {\bf130}, 104508 (2009).


\bibitem{Scriven09b}

E. Scriven and B. J. Powell, Phys. Rev. B {\bf 80}, 205107 (2009).


\bibitem{Fisher12}

H.-C Jiang {\it et. al.}, Nature {\bf 493}, 39 (2012).

















\end{thebibliography}
\end{document}